\documentclass[modern]{aastex63}
\usepackage{mdframed}
\usepackage{physics}
\usepackage[colorinlistoftodos]{todonotes}
\usepackage{verbatim}
\usepackage{cprotect}
\usepackage{multirow, booktabs}
\usepackage{listings}
\usepackage{xspace}
\usepackage{ulem}

\newcommand{\version}{0.8.0}
\newcommand{\npulsars}{45}
\newcommand{\ntoas}{12433}
\newcommand{\dof}{12307}
\newcommand{\pchi}{12368.10}
\newcommand{\pwrmspre}{0.944}
\newcommand{\pwrmspost}{0.944}
\newcommand{\tpversion}{13.101}

\newcommand{\tpchi}{12368.46}
\newcommand{\tptversion}{2019.01.1}
\newcommand{\tptchi}{12265.16}
\newcommand{\tptwrms}{0.944}
\newcommand{\testcover}{$58.05\%$}
\newcommand{\tp}{\textsc{Tempo}\xspace}
\newcommand{\pint}{\textsc{PINT}\xspace}
\newcommand{\tpt}{\textsc{Tempo2}\xspace}
\newcommand{\TP}{\textsc{Tempo}/\textsc{Tempo2}\xspace}
\renewcommand{\NG}{NANOGrav\xspace}
\newcommand{\GW}{gravitational wave\xspace}
\newcommand{\GWs}{gravitational waves\xspace}
\newcommand{\GWbar}{gravitational-wave\xspace}

\providecommand{\e}[1]{\ensuremath{\times 10^{#1}}}
\newcommand{\se}[2]{\ensuremath{{#1}\times 10^{#2}}}

\shorttitle{\pint}
\shortauthors{Luo et al.}

\begin{document}

\title{\pint: A Modern Software Package for Pulsar Timing}

\author[0000-0001-5373-5914]{Jing Luo}
\affiliation{University of Texas at San Antonio, San Antonio, TX 78249, USA}
\affiliation{Center for Advanced Radio Astronomy, University of Texas Rio Grande Valley, Brownsville, TX 78520, USA}
\affiliation{Canadian Institute for Theoretical Astrophysics, University of Toronto, Toronto, ON M5S3H8, CANADA}

\author[0000-0001-5799-9714]{Scott Ransom}
\affiliation{National Radio Astronomy Observatory, 520 Edgemont Road, Charlottesville, VA 22903, USA}

\author{Paul Demorest}
\affiliation{National Radio Astronomy Observatory, P.O. Box O, Socorro, NM 87801, USA}

\author[0000-0002-5297-5278]{Paul S. Ray}
\affiliation{U.S. Naval Research Laboratory, Washington, DC 20375-5352 USA}

\author{Anne Archibald}
\affiliation{Newcastle University, NE1 7RU, United Kingdom}

\author[0000-0002-0893-4073]{Matthew Kerr}
\affiliation{U.S. Naval Research Laboratory, Washington, DC 20375-5352 USA}

\author{Ross J. Jennings}
\affiliation{Department of Astronomy, Cornell University, Ithaca, NY 14853, USA}

\author[0000-0002-4576-9337]{Matteo Bachetti}
\affiliation{INAF-Osservatorio Astronomico di Cagliari, via della Scienza 5, I-09047 Selargius, Italy}

\author[0000-0002-6428-2620]{Rutger van Haasteren}
\altaffiliation{Currently employed at Microsoft Corporation}
\affiliation{Jet Propulsion Laboratory, California Institute of Technology, Pasadena CA 91109}

\author[0000-0001-5438-540X]{Chloe A. Champagne}
\altaffiliation{NREIP Intern at U.S. Naval Research Laboratory}
\affiliation{Vanderbilt University, Nashville, TN 37235, USA}
\author{Jonathan Colen}
\affiliation{Naval Research Enterprise Internship Program (NREIP), resident at U.S. Naval Research Laboratory, Washington, DC 20375-5352 USA}

\author[0000-0002-2099-0254]{Camryn Phillips}
\affiliation{University of Virginia, Charlottesville, VA 22903, USA}

\author[0000-0002-2917-9257]{Josef Zimmerman}
\affiliation{University of Virginia, Charlottesville, VA 22903, USA}

\author[0000-0002-7261-594X]{Kevin Stovall}
\affiliation{National Radio Astronomy Observatory, P.O. Box O, Socorro, NM 87801, USA}

\author[0000-0003-0721-651X]{Michael T. Lam}
\affiliation{School of Physics and Astronomy, Rochester Institute of Technology, Rochester, NY 14623, USA}
\affiliation{Laboratory for Multiwavelength Astrophysics, Rochester Institute of Technology, Rochester, NY 14623, USA}

\author{Fredrick A. Jenet}
\affiliation{Center for Advanced Radio Astronomy, University of Texas Rio Grande Valley, Brownsville, TX, USA}
\affiliation{University of Texas at San Antonio, San Antonio, TX, USA}

\correspondingauthor{Jing Luo}
\email{luojing1211@gmail.com}

\begin{abstract}
 Over the past few decades, the measurement precision of some pulsar-timing experiments has advanced from $\sim$10\,$\mu$s to $\sim$10\,ns, revealing many subtle phenomena.  Such high precision demands both careful data handling and sophisticated timing models to avoid systematic error.  To achieve these goals, we present \pint (\textbf{P}INT \textbf{I}s \textbf{N}ot \textsc{\textbf{T}empo3}), a high-precision \lstinline{Python} pulsar timing data analysis package, which is hosted on GitHub and available on Python Package Index (PyPI) as \lstinline{pint-pulsar}. \pint is well-tested, validated, object-oriented, and modular, enabling interactive data analysis and providing an extensible and flexible development platform for timing applications. It utilizes well-debugged public \lstinline{Python} packages (e.g.,~the \textsc{NumPy} and \textsc{Astropy} libraries) and modern software development schemes~(e.g., version control and efficient development with \lstinline{git} and GitHub) and a continually expanding test suite for improved reliability, accuracy, and reproducibility. \pint is developed and implemented without referring to, copying, or transcribing the code from other traditional pulsar timing software packages (e.g.,~\TP) and therefore provides a robust tool for cross-checking timing analyses and simulating pulse arrival times. In this paper, we describe the design, usage, and validation of \pint, and we compare timing results between it and \tp and \tpt.
\end{abstract}


\section{Introduction} \label{sec:intro}

Since their discovery in 1967~\citep{1968Natur.217..709H}, the study of pulsars has yielded major advances in a wide range of physics and astrophysical problems. Pulsars are natural laboratories for studying extreme magnetic fields ~\citep{2008Sci...319.1802G,Makishima2016}, equations-of-state of dense matter~\citep{2010Natur.467.1081D,Antoniadis+2013,Cromartie+2020}, and theories of gravity~\citep{2018Natur.559...73A, 2006Sci...314...97K, 1991ApJ...366..501D}. The most powerful aspect of pulsars is the regularity of their pulses, enabling their use as clocks spread throughout our galaxy. Pulsar timing is the technique by which observed pulse arrival times are compared to predicted arrival times based on a physical model of the pulsar signal and its propagation to the observatory. This technique can be used to study both the pulsar itself as well as the effects of binary companions (where applicable), the interstellar medium ~\citep{2017ApJ...841..125J,Donner+2019}, and Galactic dynamics~\citep{2009MNRAS.395.2326K,Verbunt+2017}.

Millisecond pulsars \citep[MSPs;][]{1982Natur.300..615B} have undergone a period of accretion from a companion star, the end result of which is often a very stable, fast-spinning pulsar (spin period $\lesssim$ 10\,ms).
Via the long-term observations of high-quality MSPs, whose pulse arrival times can be measured to better than 1\,$\mu$s, the pulsar timing technique can achieve the precision required for detecting ultra-low frequency ($\sim$ $10^{-9}$ Hz) \GWs ~\citep{fb1990,2016ApJ...819L...6T}, whose realistic astrophysical amplitudes in pulsar timing residuals will be of the order of 10\,ns.
The North American Nanohertz Observatory for Gravitational Waves~\citep[NANOGrav;][]{2013CQGra..30v4008M} is an ongoing effort to detect nanohertz frequency \GWs by monitoring a set of well-timed MSPs using the 305-m William E. Gordon Telescope (Arecibo)
of Arecibo Observatory\footnote{\href{https://naic.edu/index_scientific.php}{https://naic.edu/index\_scientific.php}} and the 100-m Robert C. Byrd Green Bank Telescope (GBT) of the Green Bank Observatory\footnote{\href{https://greenbankobservatory.org/telescopes/gbt/}{https://greenbankobservatory.org/telescopes/gbt/}}.
The international effort of pulsar timing for \GWs is under the International Pulsar Timing Array~\citep[IPTA;][]{ipta} consortium, comprising \NG, the European Pulsar Timing Array~\citep[EPTA;][]{epta}, the Parkes Pulsar Timing Array~\citep[PPTA;][]{PPTA}, and recent efforts started in South Africa \citep[MeerTime;][]{MeerTime}, India \citep[InPTA;][]{InPTA}, and China \citep[CPTA;][]{CPTA}.  

Pulsar timing for \GWs requires a good understanding of many astrophysical processes that impact the pulse times-of-arrival~(TOAs), including the pulsar system dynamics~(e.g., pulsar spin, pulsar system motion, and proper motion, etc.), solar system dynamics~(e.g., motions of the Earth and planets), and the effects of the interstellar medium~(e.g., dispersion and scintillation). 
Timing is done for each pulsar by creating a mathematical model for these effects, and then refining this model via fitting to the observed TOAs. 
For decades, the vast majority of radio pulsar timing has been accomplished using one of two major software packages: \tp\footnote{\href{http://tempo.sourceforge.net}{http://tempo.sourceforge.net}} and \tpt\footnote{\href{https://bitbucket.org/psrsoft/tempo2}{https://bitbucket.org/psrsoft/tempo2}}~\citep{tempo2}.

A robust future detection of \GWs using pulsar timing will require results to be verified with independent software packages.  However, the underlying \tpt code largely consists of \tp Fortran-based algorithms, updated to use C.
Due to the similarities in these two codes, it is necessary to develop an independent pulsar timing package for cross-checking.
The growth of computational power has allowed for high-level scripting languages, such as \lstinline{Python}, to become more popular in astronomical applications.
\lstinline{Python} has many advantages including brevity, modularity, ease of documentation, robust testing, ease of code re-use, and a large community developing powerful open-source libraries for a wide range of applications.
These features considerably improve the speed of development and the code's extensiblity, allowing us to break or extend the limitations of traditional timing software. For instance, in order to add an external high-precision orbit integrator for the pulsar triple system~\citep{triplesystem} or use a spline-based model to handle timing noise~\citep{dkg09} it was necessary to circumvent large parts of \TP or abandon them entirely, while \pint is designed to permit use of only the relevant parts or easy addition of user-written components. 
In addition, modern version control and distributed development environments like \lstinline{git} and GitHub\footnote{\href{https://git-scm.com/}{https://git-scm.com/}, \href{https://github.com/}{https://github.com/}} have facilitated community contributions that have greatly increased the pace of development and sped the adoption of these packages by the astronomical community.
Motivated by the reasons mentioned above, a new pulsar timing software project, \pint, was launched in 2013 by the \NG collaboration. 

The \pint project\footnote{Available at \url{https://github.com/nanograv/PINT} and \url{https://pypi.org/project/pint-pulsar/}} has developed a \TP-independent \lstinline{Python} toolkit --- the \pint software package --- for high-precision pulsar timing analysis to precisions of $\sim$1\,ns\footnote{For most machines on which \pint\ will be run, that $\sim$1\,ns level of precision is set by the hardware supported 80-bit floating point numbers used for many of the time-based calculations.}, and including known physical effects with timing amplitudes of $\sim$1\,ns or greater.
The \pint software package follows modern software development schemes and practices: object oriented design, modularized classes and components, a documented programming interface, and an automated test suite that is run after every change.
A major feature of the \pint package is the use of well debugged libraries
such as 
\textsc{NumPy}\footnote{\href{http://www.numpy.org/}{http://www.numpy.org/}} \citep{numpy1, numpy2}, 
\textsc{SciPy}\footnote{\href{https://www.scipy.org/}{https://www.scipy.org/}} \citep{scipy},
and \textsc{Astropy}\footnote{\href{http://www.astropy.org/}{http://www.astropy.org/}} \citep{astropy1, astropy2}. 
Because of their large active user and developer community, such packages are improved frequently and tested thoroughly. 
The dependency on such packages increases development and maintenance efficiency. 
Conversely, a key goal of \pint is that it be usable as a library itself, so key functions from \pint can be used in other pulsar-timing-related applications (for example, correcting light travel time delays in high-energy photon arrival times).

In this paper, we present an overview of the \pint pulsar timing analysis package---the full software documentation is available online\footnote{\url{https://nanograv-pint.readthedocs.io/en/latest/}}. In \S\ref{sec:timing}, we give a brief background of pulsar timing methodology. We then describe the \pint software package, including its setup, code architecture and key modules, in \S\ref{sec:pint}. In \S\ref{sec:compare}, we present one example of a \pint analysis and compare it with \TP. The tests and maintenance procedures are discussed in \S\ref{sec:unittest_mainten}.  We also introduce common use cases and their command-line scripts in \S\ref{sec:use_cases}.

\section{Overview of Pulsar Timing} \label{sec:timing}
Pulsar timing refers to the process of unambiguously, and to high precision, accounting for pulse TOAs at a telescope using a relatively simple timing model. Here we give a brief overview of pulsar timing including (i) obtaining TOAs, (ii) modeling the pulse emission and propagation time, (iii) comparing the model to observed data, and (iv) improving the model.

\subsection{Measuring TOAs} \label{subsec:TOAs}
The key measurement for pulsar timing, a TOA, notionally measures the time when a fiducial point of a pulsar pulse profile reaches an observer. Normally, these measurements are actually made on the coherent average of many pulses, the \emph{folded pulse profile}, both to increase the signal-to-noise ratio and to mitigate the effects of pulse-to-pulse variations \citep{handbook,cd85}. 
This coherent average process, also called ``folding'', sums the pulse based on their pulse phases~(see \S\ref{subsubsec:phase} for the definition of ``pulse phase''), which are computed from the existing pulsar timing model. 
In the case of high-energy observations, such as from X-ray or $\gamma$-ray observatories, TOAs are not necessarily the focus; individual photon arrival times have their pulse phases computed and can be binned into a pulse profile \citep{rkp+11} or treated individually \citep{cp15}.

Given an observation of a pulsar, one generally compares the folded pulse profile to a known template describing the pulsar's (usually stable) pulse profile. A template-matching algorithm \citep[e.g.,][]{1992RSPTA.341..117T} permits very accurate computation of a shift, expressed in units of rotational phase from $-0.5$ to $0.5$, of the observed pulse compared to the template. Phase zero denotes perfect alignment with the template. This computed phase shift is then used to construct a TOA. This begins with the phase-zero moment (according to the ephemeris used for folding) nearest the middle of the observation span and adjusts that time by the measured phase shift multiplied by the pulse period. The TOA is thus the idealized arrival time of the phase-zero part of the template near the middle of the observation span. The TOA value itself is generally represented as a Modified Julian Day (MJD) in the Coordinated Universal Time (UTC) time system\footnote{This has known problems; see \S\ref{subsec:pintcoords}.}, as recorded by an observatory clock.  The TOAs require certain additional data, including the observatory where the TOA was recorded, an estimate for the error in the determination of the TOA, and the radio frequency at which it was recorded. Further information can also be recorded, such as the pulsar name, the signal-to-noise ratio of the measurement, the instrument with which it was recorded, \emph{et cetera}.

\pint does not provide functionality for measuring TOAs, that is left for codes specific to particular types of data. 
But, \pint can be used to compute the pulse phases for data folding or other calculations~(e.g., photon phases). For instance, it has a module to generate and interpolate the coefficients of polynomial approximations of the pulse phase (i.e., polycos).

\subsection{Modeling TOAs} \label{subsec:models}
In order to understand the physics behind the TOAs, we compare them to a timing model, which is mathematical description of (i) the rotation of the pulsar and (ii) the propagation of its pulses to the observer. The pulsar rotation is mathematically represented using rotational phase. The propagation process is modeled in terms of time delays related to the light travel time from the pulsar to the observer. In the following subsections, we describe these two parts in more detail.

\subsubsection{Rotational Phase}\label{subsubsec:phase}
Rotational phase, often referred to as simply phase, describes a pulsar's rotational status in a reference frame that is co-moving with the pulsar. One complete rotation is represented by an increase in phase of 1. As the pulsar rotates, the phase naturally increases, and is often written as $N(t)$, the cumulative phase number. In cases where the absolute pulse number is not needed or not available, the integer portion may be ignored, and a wrapping fractional phase ranging from 0 to 1 is used. There is some arbitrariness in the definition of phase zero; it is usually defined as the zero in phase of an idealized pulse profile template; this is frequently chosen to be either the highest point or center of mass of the profile, for pulsars whose profile consists of only a single component.

Since pulsars do not rotate at constant pulse frequencies, a Taylor expansion typically describes the rotational phase as:
\begin{eqnarray}\label{eq:phase_number}
N(t) = N_{0} + \nu_{0}(t-t_{0}) +  \frac{1}{2}\dot{\nu_{0}}(t - t_{0})^{2} + \frac{1}{6}\ddot{\nu_{0}}(t-t_{0})^{3} + \ldots,
\end{eqnarray}
where $N_{0}$ is the phase/pulse number at a reference epoch $t_{0}$, $\nu_{0}$ is the pulse frequency (i.e.,~the first time derivative of the phase) at $t_{0}$, and $\dot{\nu_{0}}$ and $\ddot{\nu_{0}}$ are the first and second derivatives of pulse frequency \citep[e.g.,][]{handbook}. More complicated rotational models are possible, for instance those with glitches~(a sudden change in pulse frequency; \citealt{glitch}) and glitch relaxation.

If we choose one pulse's arrival time as our reference time $t_{0}$, our model parameters are known exactly, and without noise, then the phase at other pulse arrival times $N(t_{\rm TOA})$ will be an integer value.

Practically, in order to evaluate Eqn.~\ref{eq:phase_number} we must transform our observed TOAs into the pulsar co-moving frame. 
In the next sub-section \S\ref{subsubsec:delay}, these transformations, including time scale conversions and propagation time modeling, are discussed.

\subsubsection{Pulse Delays}\label{subsubsec:delay}
The delay portion of the timing model characterizes the total pulse propagation time, determined by a variety of physical processes between the pulsar and the observer.
Given the TOA at the observatory, we can compute the pulse emission time via the total delay,
\begin{eqnarray} \label{eq:emission_time}
t_{\rm e} = t_{\rm obs} - \Delta,
\end{eqnarray}
where $t_{\rm e}$ is the pulse emission time, $t_{\rm obs}$ is  the pulse observation time and $\Delta$ represents the total delay, from a wide variety of causes.
The total delay,
\begin{eqnarray} \label{eq:total_delay}
\Delta = \Delta_{\rm A} + \Delta_{\rm R \odot} + \Delta_{\rm E \odot} + \Delta_{\rm S \odot}  + \Delta_{\rm SB} + \Delta_{\rm fd} + \Delta_{\rm binary} + \ldots,
\end{eqnarray}
where we have listed the most common delays in the timing process~\citep[e.g.,][]{handbook}.
The first term $\Delta_{\rm A}$ represents the delay caused by the ``hydrostatic'' atmospheric effects of topocentric observations, modeled as the product of the delay at zenith (\citealt{davis1985}) and an azimuthally symmetric function that maps the delay onto any other position in the sky (\citealt{niell1996}).
The next three terms, $\Delta_{\rm R \odot}$, $\Delta_{\rm E \odot}$, and $\Delta_{\rm S \odot}$. are the Solar System geometric or R{\o}mer delay, Einstein delay (comprised of gravitational redshift and time dilation; \citealt{einsteindelay}), and Solar System Shapiro delay (due to the gravitational perturbation of the light-path; \citealt{shapirodelay}).
Although the Shapiro delay term formally includes contributions from all Solar System bodies, we normally only include those from the Sun and major planets~(i.e., time delays bigger than 1\,ns; \citealt{tempo2}).  
The $\Delta_{\rm SB}$ term gives the light travel time from the pulsar system to the solar system.
Its initial value, which is a very large quantity, can be absorbed in the phase calculation since a phase is computed relative to a reference epoch (see below). 
The time-dependent part of this delay due to relative motion is separated into delays that vary due to transverse and radial motion of the pulsar.
The former is modeled as the proper motions via the solar system R{\o}mer delay; however, the radial component effect is generally hard to distinguish from the pulse period derivative.
The $\Delta_{\rm fd}$ term includes a variety of radio-frequency-dependent time delays, such as the dispersion delay caused by the ionized interstellar and interplanetary media.
The last term, $\Delta_{\rm binary}$, includes the pulsar system's R{\o}mer, Einstein\footnote{This ``Einstein delay'' is not actually a delay; instead it is the cumulative effect of gravitational and special-relativistic time dilation on the pulsar. In normal pulsar work the units of time for the pulsar are rescaled so that the mean time dilation is zero and the ``Einstein delay'' oscillates around zero.\label{foot:ed}}, and Shapiro delays.
The pulsar R{\o}mer delay is controlled by the position of the pulsar at the moment of pulse emission, rather than the moment of pulse arrival at the Solar System Barycenter. 
Thus, $\Delta_{\rm binary}$ needs to be evaluated at a time that needs $\Delta_{\rm binary}$ itself as input; older timing models incorporate an approximate solution to this inversion problem in their formulas~\citep{DD_model}, while more modern ones solve it directly by root-finding~\citep{triplesystem}.
These delay terms' typical range of values are summarized in the \citet{tempo2} Table 2.

Given the transformation from pulse observed time $t_{\rm obs}$ to pulse emission time (ignoring a constant pulsar system Einstein delay, see footnote~\ref{foot:ed})

\begin{eqnarray}
N(t_{\rm obs}) = N_{0} + \nu_{0}(t_{\rm obs} - \Delta -t_{0}) +  \frac{1}{2}\dot{\nu_{0}}(t_{\rm obs} - \Delta - t_{0})^{2} + \frac{1}{6}\ddot{\nu_{0}}( t_{\rm obs} - \Delta -t_{0})^{3} + \ldots,
\end{eqnarray}
The computed phases are described relative to a reference phase $N_{0}$ at the reference time $t_{0}$. 
In practice, $N_{0}$ is defined by specifying a moment at which the phase is zero ($N = 0$). This moment is specified in the reference frame by a reference MJD, observatory site, and radio frequency (often denoted by the parameters \texttt{TZRMJD}, \texttt{TZRSITE}, \texttt{TZRFRQ}), as was done in \TP.  \texttt{TZRMJD} is treated as a hypothetical arrival time measurement, in the timescale of the observatory clock.  To transform that time to other timescales, standard clock corrections need to be applied as per any other TOA (see \S\ref{subsec:pintcoords}). The resulting phases are used in the process of refining the timing model. Currently, if \texttt{TZRMJD} is not specified,
the phase of the first TOA in the TOAs table is defined to be zero.

\subsection{Comparing model to the data} \label{subsec:compare_model}
In order to improve the accuracy of a timing model, the timing residuals, defined as the differences between the observed TOAs and the TOAs predicted by the given timing model, are introduced,
\begin{eqnarray}\label{eq:time_res_def}
R_{\rm time} \equiv t_{\rm obs} - t_{\rm model} \label{eq:res_def}
\end{eqnarray}
Because of the periodic nature of the pulsar's signal, the residuals thus obtained are known only modulo one rotation of the pulsar --- that is, \emph{a priori} we do not know the integer number of rotations between two pulse arrival time measurements. In an established pulsar timing program, as described in \S\ref{subsubsec:phase}, our estimated model is generally accurate enough that the predicted TOAs will differ from the observed TOAs by less than one pulse period. That is, a sufficiently good model allows us to infer the exact number of rotations between any two observations. When the model is insufficient, perhaps because we are observing a new pulsar, or there has been a long gap in observations, or a glitch has occurred, the uncertain number of rotations between observations can make the task of finding or improving a timing solution a highly discontinuous and difficult optimization problem. Traditionally this has been addressed by hand, with users introducing turn-number guidance into the TOA data files, iteratively working with larger and larger subsets of observations until a satisfactory ``phase-connected'' solution has been found. Automated tools for phase connection have been implemented~\citep{fr18}. Alternatively, if precise rotation numbers have been inferred for the TOAs, these can be coded into the input files, reducing or removing the discontinuous nature of the fitting problem. 

Multiplying the time residual by the pulse frequency, we can write the residuals in terms of phase number:
\begin{eqnarray}\label{eq:phase_res}
R_{\rm phase} = N( t_{\rm obs})- N_{i}(t_{\rm obs})
\end{eqnarray}
where $N_{i}$ is the inferred integer phase number at $t_{\rm obs}$.
In terms of phase residual, the time residual can be also written as:
\begin{eqnarray}\label{eq:time_res}
R_{\rm time} &=& R_{\rm phase}/\nu(t_{\rm obs}),
\end{eqnarray}
where $\nu(t_{\rm obs})$ is the apparent pulsar pulse frequency at $t_{\rm obs}$ in the frame of the observer.
Traditionally, $\nu_{0}$, the pulsar pulse frequency at reference time $t_{0}$, has been used for this scaling and for many pulsars the error is negligible, but \pint implements this more correct time residual calculation.
From the residuals, the current timing model can be updated by using a variety of fitting methods.
Because of the issue of phase connection, pulsar timing is generally carried out in an iterative way: an approximate model is successively updated as new data becomes available or as more complex models are applied. 
In the each iteration, the previous post-fit timing model is treated as the input model and gets updated by tuning the parameter values or using new models~\citep{handbook}.

For traditional \GW detection projects, the residuals generated by a good deterministic timing model are the starting point of analyses~\citep{1979ApJ...234.1100D}. 
\citet{1983ApJ...265L..39H} describe the contribution of an isotropic \GW background on correlations in the timing residuals from an array of well-timed pulsars, that is, a Pulsar Timing Array~\citep[PTA;][]{PTA}.
A main objective for the \pint package is to provide high-quality timing and software tools for this type of analysis.
Currently, \pint can be used by  \NG's gravitational wave analysis package, the Enhanced Numerical Toolbox Enabling a Robust PulsaR Inference SuitE~(\textsc{ENTERPRISE})\footnote{\href{https://github.com/nanograv/enterprise}{https://github.com/nanograv/enterprise}}.  
In addition, \pint provides analytical derivatives of the phase with respect to most timing parameters and the capability to use numerical derivatives (i.e., finite differences) for all timing parameters (see \S\ref{subsec:models}).
Many \GWbar analyses~\citep[for example][]{vH+09} are able to use the derivatives of the residuals with respect to the timing model parameters as a more efficient proxy for the full timing model that permits analytical marginalization.

\section{pint} \label{sec:pint}

\begin{figure}[t]
\plotone{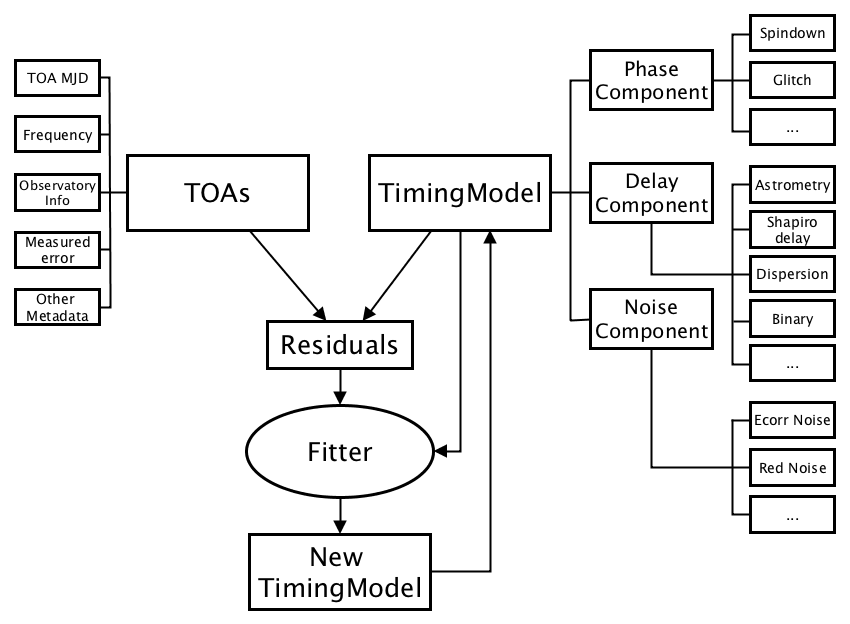}
\caption{\pint operational model.  This is a rough model as to how \pint is designed and implemented as well as how it is used for timing a pulsar. Lines without arrows indicate that the object in question contains the data; arrows indicate that results computed in one object are passed to the other.  The TOAs and timing models are kept as independent as possible and only interact through other parts of \pint functionality, such as creating residuals and fitting models to data. \label{fig:code_arch}}
\end{figure}

\pint is a \lstinline{Python} library and a set of executable scripts, compatible with \lstinline{Python} 3.6 or greater\footnote{Support for \lstinline{Python} 2.7 was dropped in 2020, in conjunction with many other astronomical \lstinline{Python} packages (see \url{http://python3statement.org})}. In this section we introduce the \pint software package version \version{} and provide code examples. The operational model of \pint is illustrated in Figure \ref{fig:code_arch}. In the following subsections, the fundamental assumptions, including the coordinate definitions and the treatment of time, are discussed first. Code architecture details and the basic application programming interface (API) of the major modules are presented afterward.

\subsection{\pint Coordinates and Time} \label{subsec:pintcoords}

As discussed in \S\ref{subsubsec:phase}, the description of the pulsar signal is relatively simple in a nearly-inertial frame, such as that of the SSB.  As with most other timing packages, \pint uses the SSB as its reference frame for pulsar timing models.  Given the design of \pint, if other reference frames were required, for instance that of the pulsar, they could be added in a relatively straightforward manner.

NASA's Jet Propulsion Laboratory (JPL) has adopted the International Celestial Reference System (ICRS) J2000 reference frame, as the base coordinate system for all of their solar system ephemeris calculations ~\citep{2014IPNPR.196C...1F}. Therefore, since \pint uses the JPL ephemerides, all internal \pint calculations are performed in this coordinate system. A pulsar's position and velocity are generally specified by astrometric parameters (e.g., Right Ascension, Declination, and proper motions in the ICRS frame) as part of a timing model. The observatories' positions and velocities are tabulated by the \pint \lstinline{observatory} module which is discussed in \S\ref{subsubsec:observatory}.  Coordinate transformations are performed using \textsc{Astropy} routines whose algorithms are provided via the Essential Routines for Fundamental Astronomy package~(\textsc{ERFA}), a re-branding of the Standards Of Fundamental Astronomy (\textsc{SOFA}) library\footnote{\href{http://www.iausofa.org/} {http://www.iausofa.org/}}~\citep {2013A&A...558A..33A}.

\pint assumes the TOAs it reads to be MJD values in the timescale of the observatory where they were recorded (the observatory timescales are handled in the \lstinline{observatory} module, see \S\ref{subsubsec:observatory}), although \pint can also accept TOAs in other ``special'' reference frames such as those at the SSB or at the geocenter.  To store these MJDs at the required numerical precision of $\sim$1\,ns, \pint uses the \lstinline{astropy.time.Time} object\footnote{\href{http://docs.astropy.org/en/stable/time/}{http://docs.astropy.org/en/stable/time/}}, where two 64-bit floats represent the integer and fractional parts of each MJD. Since there is no standard way of representing UTC times on leap days as normal MJDs,\footnote{The precision timing community knows well that using the MJD format for UTC times is fraught with peril. There is no unique way to assign MJDs to times during days with leap seconds, and MJD1$-$MJD2 does not correctly give the time interval between two times, because of possible leap seconds between MJD1 and MJD2. Nevertheless, MJDs are commonly used for UTC times in many places.}  \pint follows \tp and \tpt in defining a custom time format called \lstinline{pulsar_mjd}, in which the integer part is the normal integer MJD and the fractional part is the seconds of the day divided by 86400. The means that MJDs `tick' at a constant rate, but there is no representation for a time during a leap second, and therefore no way to represent a TOA during that time.

In order to convert TOAs to Barycentric Dynamical Time~(TDB), a sequence of clock corrections has to be applied on the TOAs. The raw TOAs are typically referenced to an observatory clock, often a GPS-disciplined rubidium clock or hydrogen maser. This timescale is denoted as UTC(obs), where ``obs'' is the name of the observatory.  \pint applies the usually-known local clock corrections to convert UTC(obs) to UTC(GPS), a timescale maintained by the U.S. Naval Observatory~(USNO).  Those corrections use either \tp or \tpt format clock files, which are obtained from observatories by various means and must be kept up-to-date. By default, \pint uses the set of \tp-format clock files distributed with \pint in \lstinline{src/pint/datafiles}. If needed, \pint is also able to read the clock correction files from \TP clock directories.  A further correction can be applied to convert UTC(GPS) to the standard UTC, maintained by the International Bureau of Weights and Measurements (BIPM), using the \tpt-format \lstinline{gps2utc.clk} file (which must also be kept up-to-date) in \lstinline{pint/datafiles}.  Those corrections are derived from BIPM Circular T\footnote{\url{https://www.bipm.org/en/bipm-services/timescales/time-ftp/Circular-T.html}}. Whether this correction is applied can be controlled via the \lstinline{observatory} API, which is discussed in \S\ref{subsubsec:observatory}.

UTC is converted to International Atomic Time (TAI; using \textsc{Astropy}) by adding an integer number of leap seconds, and then to Terrestrial Time (TT, also known as Terrestrial Dynamical Time, or TDT), which ticks at the same rate as TAI and UTC, but for reasons of continuity has an offset. A TT day has a duration 86400 seconds on the geoid and is the independent argument of apparent geocentric ephemerides.  The most common realization of TT is TT(TAI), which is defined as: TT(TAI) = TAI + 32.184~seconds. However, \pint can also use TT(BIPM), which is a more accurate realization of TT published by the BIPM. In \pint, this clock correction is read from the \tpt-style clock file \lstinline{pint/datafiles/tai2tt_bipm2015.clk} (or an alternative file based on the approximately annual publication of the BIPM timescales).  Whether this correction is enabled is controlled by the \lstinline{include_bipm} argument to \lstinline{pint.observatory.get_observatory()}, and if it is, the version of TT(BIPM) can be selected by the \lstinline{bipm_version} argument.

Finally, times are converted from TT to a barycentric time. There are two such time systems in common use. Traditionally, pulsar timing has been done using TDB, which is the independent variable of the JPL planetary ephemerides~\citep{tdbvstcb}. The alternative is Barycentric Coordinate Time~(TCB), which is the preferred timescale according to the International Astronomical Union~(IAU). TCB is a relativistic coordinate time and the modern definition of TDB is a linear scaling of TCB~(IAU Resolution 3 of 2006\footnote{\url{https://www.iau.org/administration/resolutions/ga2006/}}).  The tick rates of the two differ by about a part in $10^8$, so the value of model parameters which have a time component in the unit are different depending on the choice of barycentric timescale. Currently, \tp and \pint only support TDB, while \tpt uses TCB as its default but allows the choice of TDB for compatibility. In the future, \pint will be extended to support TCB.

\begin{figure}[htb!]
\plotone{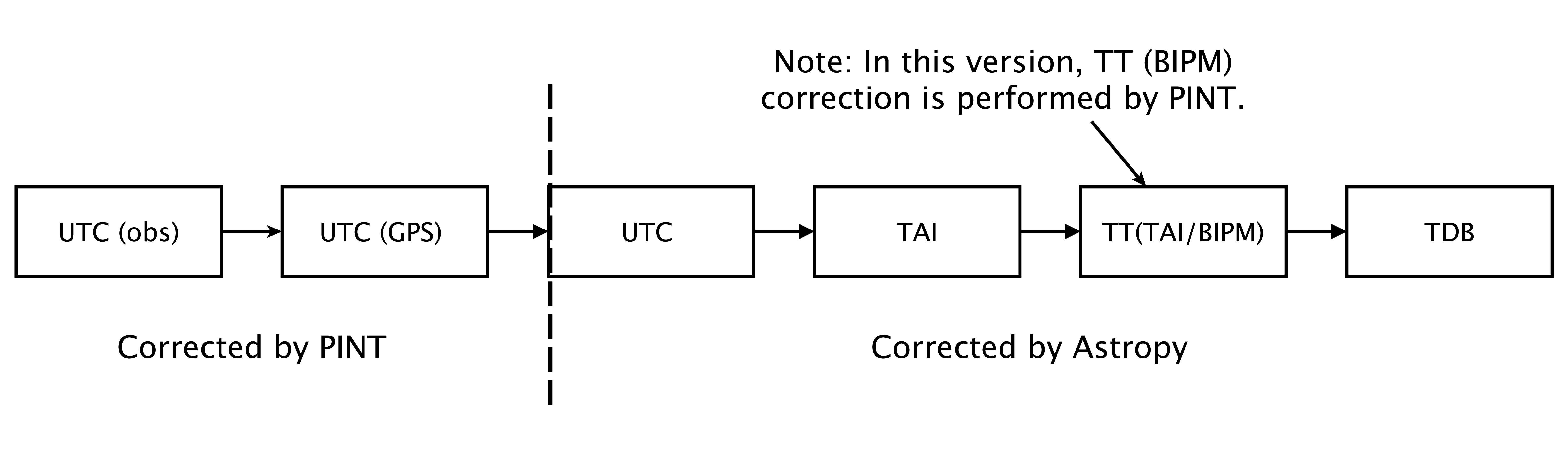}
\caption{\pint converts TOAs from the observatory local time UTC(obs) to TDB following the steps illustrated. \pint handles the conversion from UTC(obs) to UTC(GPS) and the TT(BIPM) correction. The other part of clock corrections are performed by \textsc{Astropy}. \label{fig:Time_convert}}
\end{figure}

In \pint, the default conversion from TT to TDB is handled by \textsc{Astropy}, which uses the SOFA library
to perform the conversion. The difference TDB$-$TT is quasi-periodic, dominated by an annual term of amplitude 1.7\,ms. The SOFA routines implement an approximation to this function
using over 800 terms from \citet{FB90} and include a location-dependent correction.
\pint also provides the infrastructure to incorporate other types of TT$-$TDB corrections (e.g., numerical TT$-$TDB difference provided by JPL ephemerides or the IF99 method; see \citealt{if99}).
The complete \pint clock correction chain is illustrated in Figure \ref{fig:Time_convert}.

Several of the clock corrections are based on published or measured data provided
by observatories or international organizations. Section \S\ref{subsubsec:externaldata} describes
the scheme \pint uses for reading and updating these external data sets.

Note that clock corrections as described here are independent of corrections for light travel time: although the times at the end of this process are in TDB, they have not been corrected for light travel time across the solar system and are therefore not what pulsar astronomers conventionally call ``barycentered''. That process happens later since the correction depends on astrometric parameters from the timing model and a solar system ephemeris, not just the TOAs themselves.

\subsection{\pint code architecture}\label{subsec:code architecture}
\pint is designed to be highly modular. 
According to the pulsar timing procedures introduced in \S\ref{sec:timing}, \pint organizes its code in four major independent modules: \lstinline{pint.toa}, \lstinline{pint.models}, \lstinline{pint.residuals}, and \lstinline{pint.fitter}.

The \lstinline{pint.toa} module provides the container classes used to store and manipulate TOAs and their corresponding metadata, while \lstinline{pint.models} contains the classes that implement the various timing models to predict TOAs.  The \lstinline{pint.fitter} module provides classes which vary timing model parameters to optimally fit the TOAs. Typically such a comparison between the TOAs and timing model occurs through the use of the \lstinline{pint.residuals} module.

Each of these modules provides public interface classes for common usages.
The classes \lstinline{TOAs}, \lstinline{TimingModel}, and \lstinline{Residuals} are used to interface with the modules \lstinline{pint.toa}, \lstinline{pint.models}, and \lstinline{pint.residuals}, respectively.
These interface classes can be initialized independently, allowing one to, for instance, analyze details of a pulsar's timing model without having TOAs from the pulsar. This flexibility is one of the key innovations of the \pint package.
The interface to the \lstinline{pint.fitter} module depends on the chosen fitting method (e.g., the  \lstinline{WLSFitter} class for a weighted least squares fit versus the \lstinline{GLSFitter} class for generalized least squares), but all fitter classes require instances of both \lstinline{TOAs} and \lstinline{TimingModel}, which are compared internally using \lstinline{Residuals}. Table \ref{tab:key modules} lists the frequently used modules in \pint.

\floattable
\begin{deluxetable}{c c c }
\tablecaption{\pint common modules.\label{tab:key modules}}
\tablewidth{0pt}
\tablehead{
\colhead{Module Name} & \colhead{Provides} & \colhead{Reference Section} 
}
\tablecolumns{4}
\startdata
  \lstinline{toa} &TOA\tablenotemark{a} container and API & \ref{subsec:TOAs} \\
  \lstinline{observatory} &Observatory's position, velocity and clock corrections & \ref{subsubsec:observatory}\\
  \lstinline{models} & Timing model API and built-in model components & \ref{subsec:models}  \\
  \lstinline{residual} & Residual container and API & \ref{subsec:residual} \\
  \lstinline{fitter} & Fitter API and built-in fitting algorithms & \ref{subsec:fitter} \\
  \lstinline{pintk} & \pint Graphical user Interface & \ref{sec:use_cases} \\
  \lstinline{scripts} &Commonly used command-line scripts& \ref{sec:use_cases}\\
\enddata
\tablenotetext{a}{Time of Arrival}
\end{deluxetable}

\begin{figure}[ht!]
\begin{mdframed}
\plotone{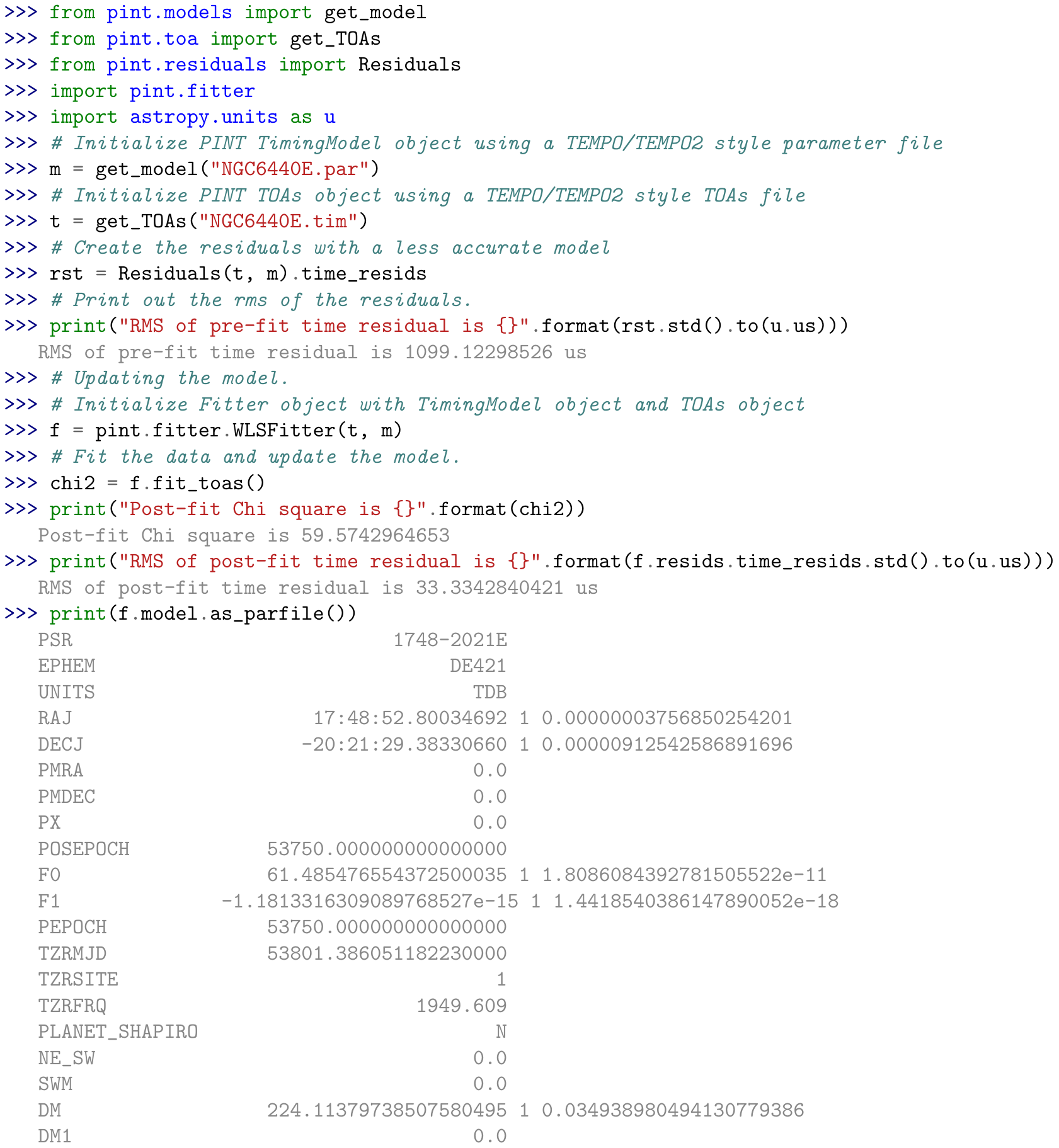}
\end{mdframed}
\caption{Code example showing \pint being used like \tp to  update an existing pulsar timing model using observed TOAs\label{fig:example_update_model}.}
\end{figure}

One of the most common uses of \pint is to mirror the standard \tp functionality of updating existing timing models using newly observed data.
All four modules must be used together in order to achieve this functionality.
The code example in Figure~\ref{fig:example_update_model} demonstrates how to use \pint as a substitute for \TP, and the four primary \pint classes or class types: \lstinline{TOAs}, \lstinline{TimingModel}, \lstinline{resids}, and the fitting classes in \lstinline{pint.fitter} work together following the operation model in Figure \ref{fig:code_arch}.

In the following sections, these four key modules and APIs will be discussed in detail.

\subsection{TOA module} \label{subsec:TOAmodule}
As introduced above, the \lstinline{pint.toa} module provides the container class (\lstinline{TOAs}) and APIs for reading, processing, storing, and interacting with TOAs. However, during TOA processing, the \lstinline{pint.observatory} module also plays an essential role behind the scenes.

\subsubsection{Handling TOAs} \label{subsubsec:handling_toas}
Typically, a user will read in and preprocess TOAs using the convenience function \lstinline{toa.get_toas()} as shown in the code example in Figure~\ref{fig:example_update_model} and discussed in \S\ref{subsec:code architecture}.
The TOAs and associated metadata (e.g.,~observing frequencies, TOA errors, observatories used, etc.) are typically read from a set of text files known as \textsc{.tim} files.
Currently, \lstinline{toa.get_toas()} can read ``Princeton'', ``Parkes'', and ``\tpt'' format TOAs\footnote{\url{http://tempo.sourceforge.net/ref_man_sections/toa.txt} }.
All the TOA information is stored in the publicly accessible attribute \lstinline{TOAs.table}, which is an instance of an \lstinline{astropy.Table} object, allowing \pint to take advantage of the latter's high-level table access and manipulation capabilities.  For example, table columns and associations can be easily defined or modified, and subsets of TOAs can easily be selected or de-selected.

The \lstinline{toa.get_toas()} function processes the raw TOAs upon reading using three \lstinline{TOAs} class methods: \lstinline{apply_clock_corrections()}, \lstinline{compute_TDBs()}, and \lstinline{compute_posvels()}. These methods transform the TOAs to the TDB timescale and compute the solar system objects' positions and velocities in the ICRS J2000 coordinate system at those times.
Since the coordinate and time transformations are highly observatory dependent, these three \lstinline{TOAs} class methods are actually high-level wrappers of several detailed computations provided in the \lstinline{observatory} module, which is discussed in \S\ref{subsubsec:observatory}. The \lstinline{toa.get_toas()} method also allows the user to control the version of external data~(\S\ref{subsubsec:externaldata} discusses the external data handling scheme) used in these wrapped functions via the input arguments. Traditionally, this information are stored in the timing model parameter~(\textsc{.par}) files, which are processed by \lstinline{pint.models} module. To avoid the inconvenience, \pint{} \version{}'s \lstinline{toa.get_toas()} accepts the \lstinline{TimingModel} object, where the versions of external data are saved, as an input argument and applies them to the TOAs.
The read-in and clock-corrected TOAs are stored in the \lstinline{TOAs.table["mjd"]} column as \lstinline{astropy.time.Time} objects\footnote{The \lstinline{astropy.time.Time} object uses a pair of 64-bit floating-point numbers to represent times (integer and fractional parts of the Julian Day number) and as a result is capable of 20\,ps precision. Unfortunately few mathematical operations can be used directly on these objects.}. The use of tables allows for flexible organization and handling of TOAs, allowing users and developers the ability to quickly and efficiently index and select TOAs.
As a convenience, and with approximately the same $\sim$1\,ns precision, the TDB times in MJD format from \lstinline{compute_TDBs()} are stored in the \lstinline{TOAs.table["tdbld"]} column as a \lstinline{np.longdouble}\footnote{The type \lstinline{np.longdouble} uses the underlying C implementation's \lstinline{long double} type. On most Intel machines this is hardware-supported 80-bit floating-point packed into larger blocks of memory. The Microsoft Visual C runtime defines this type to have only 64 bits, and so PINT cannot run there. Other machines may define \lstinline{long double} to be either software or hardware supported quadruple precision or software-supported double-double precision (for example Arm64, Power9, and Power7 architectures respectively). In any case PINT will refuse to run if this data type cannot support nanosecond precision on MJDs.} array, which can be directly used in most \textsc{NumPy} and \textsc{SciPy} vector calculations.
Some intermediary results of the time transformations (e.g., TOAs in Terrestrial Time) are saved in additional \lstinline{TOAs.table} columns, allowing the user to have easy access to these results, if needed.
Observatories' positions and velocities, using \textsc{Astropy} quantities and units, in the ICRS J2000 frame are computed by the \lstinline{compute_posvels()} class method and saved in the \lstinline{TOAs.table} columns \lstinline{"ssb_obs_pos"} and \lstinline{"ssb_obs_vel"}, respectively. 
The positions of the Sun and major planets are also computed by \lstinline{compute_posvels()}, to enable solar system Shapiro delay calculations. Table \ref{tab:table info} lists the \lstinline{TOAs.table} columns after calling the \lstinline{get_toas()} function.  For efficiency, \pint can pickle\footnote{Pickling is a process that serializes a python object to a binary format that can be efficiently written to a file. \href{https://docs.python.org/3/library/pickle.html}{https://docs.python.org/3/library/pickle.html}} the TOAs and computed data for later use, if the \lstinline{usepickle} flag is enabled in \lstinline{get_toas()}. 
The performance difference between pickling and non-pickling is discussed in \S \ref{subsec:performance}.
\begin{figure}[ht!]
\begin{mdframed}
\plotone{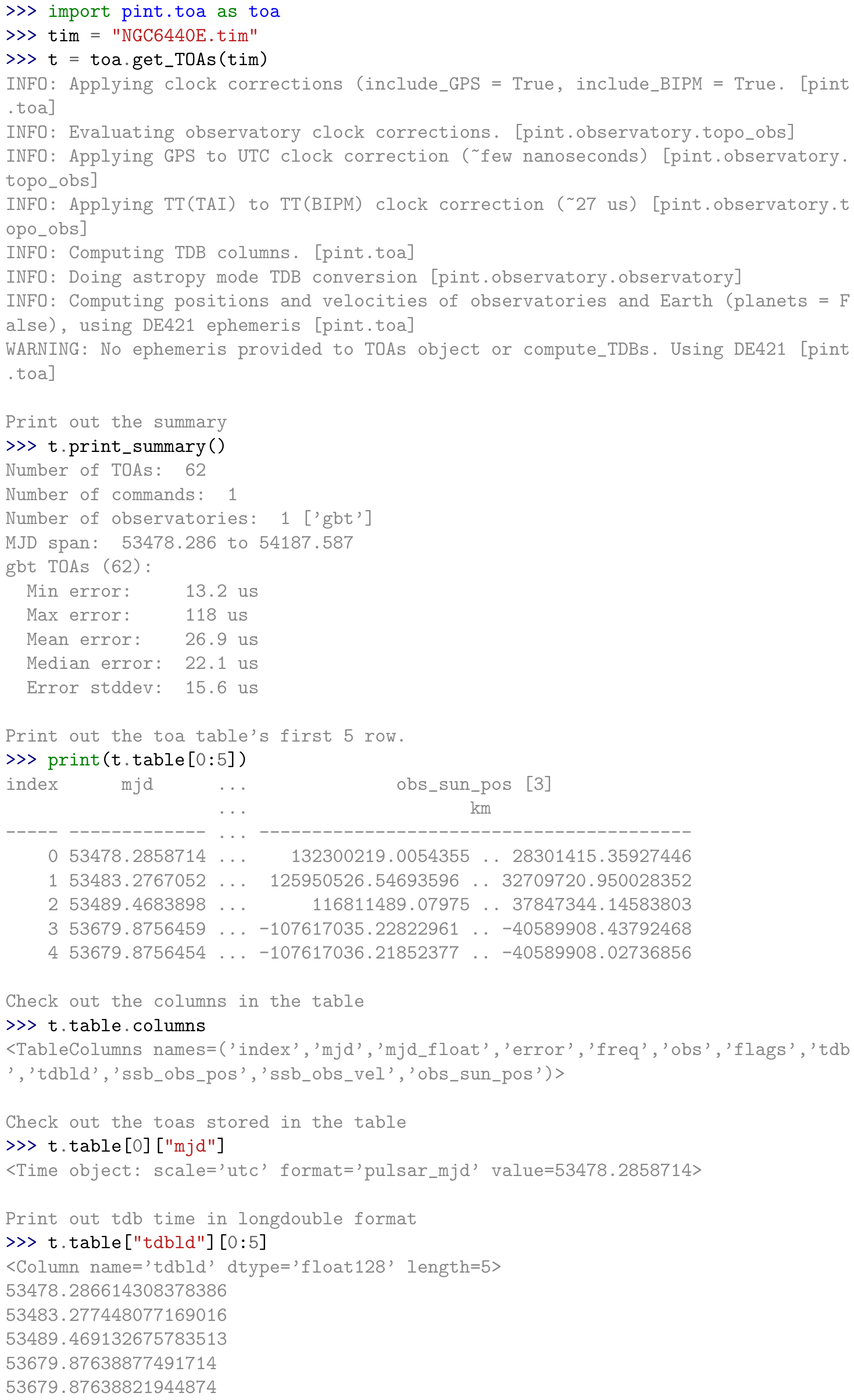}
\end{mdframed}
\caption{Code example for TOA module}\label{fig:toa_code}
\end{figure}

\floattable
\begin{deluxetable}{c c c c}
\tablecaption{Information stored in the \lstinline{TOAs.table} object.\label{tab:table info}}
\tablewidth{0pt}
\tablehead{
\colhead{Column Name} & \colhead{Descriptions} & \colhead{Data type} & \colhead{Unit}
}
\tablecolumns{4}
\startdata
   mjd & TOA\tablenotemark{a} at Observatory in UTC & \lstinline{astropy.time.Time} & MJD\\
  error& TOA error & \lstinline{np.float} & $\mu$s \\
  freq & TOA observing frequency & \lstinline{np.float} &MHz\\
  obs & Observatory name/code & \lstinline{str} &None\\
  flags & Command flags & \lstinline{dict} &None\\
  tdb & TOA in TDB\tablenotemark{b} & \lstinline{astropy.time.Time} &MJD\\
  tdbld &  TOA in TDB in long double format & \lstinline{np.longdouble} &MJD\\
  ssb\_obs\_pos & SSB\tablenotemark{c} $\rightarrow$ Observatory position vector & \lstinline{np.float} &km\\
  ssb\_obs\_vel & Observatory velocity (referenced to SSB)  &\lstinline{np.float} &km/s\\
  obs\_sun\_pos & Observatory $\rightarrow$ Sun center position vector & \lstinline{np.float} & km\\
\enddata
\tablenotetext{a}{Time of Arrival}
\tablenotetext{b}{Barycentric Dynamical Time}
\tablenotetext{c}{Solar System Barycenter}
\end{deluxetable}

\subsubsection{Handling Observatories}\label{subsubsec:observatory}
The \lstinline{observatory} module stores fundamental observatory information and provides additional coordinate and time transform functionality, for both stationary and moving observatories (i.e., satellites). The base class, \lstinline{Observatory}, provides the unified API for obtaining observatory positions and velocities, computing the clock correction values, and calculating time transformations to TDB, with the methods \lstinline{posvel()}, \lstinline{clock_corrections()}, and \lstinline{get_TDBs()}, respectively. However, as these calculations may be observatory specific, their implementations are in the various \lstinline{Observatory} subclasses. This scheme allows \pint to handle TOAs from different observatories simultaneously and clearly.

There are currently two \lstinline{observatory} subclasses, \lstinline{TopoObs} and \lstinline{SpecialLocation}. The \lstinline{TopoObs} class is implemented for stationary ground-based observatories, such as most traditional radio telescopes (e.g., Arecibo Observatory and Green Bank Observatory). Ground-based observatories follow the standard procedure of coordinate transformation and clock correction from the Earth co-rotating frame to the ICRS frame (i.e., applying the clock corrections and coordinate transformations introduced in \S\ref{subsec:pintcoords}).  Creating a \lstinline{TopoObs} object requires the observatory name, aliases (i.e., as often used on TOA lines), and coordinates under the International Terrestrial Reference Frame\footnote{\href{http://itrf.ensg.ign.fr/}{http://itrf.ensg.ign.fr/}}~\citep[ITRF;][]{ITRF}.

In contrast, the \lstinline{SpecialLocation} class is designed to implement the observatories that are not in a fixed location co-rotating with the Earth, such as the imaginary solar system barycenter (SSB) ``observatory'' or an Earth-centered ``observatory'' (i.e., the geocenter). Another use case for the \lstinline{SpecialLocation} class are space-based observatories such as \emph{Fermi}~\citep{fermi} and \emph{NICER}~\citep{NICER}, where orbital information or other spacecraft flight data is required rather than ITRF coordinates.  Detailed and observatory-specific calculations are provided by individual \lstinline{Observatory} objects, whereas the \lstinline{SpecialLocation} class implements only the high-level APIs for these calculations. 

In the current \pint version, many observatories, both real and imaginary (like the geocenter and SSB) are pre-defined in the \lstinline{observatory} module.  Most users will create an observatory instance with the convenience function \lstinline{get_observatory()}, which takes the observatory string name or \tp style observatory code as an input argument. Special position/velocity or time transformation algorithms and their required external data sets or versions can be selected with optional arguments (e.g., the \lstinline{include_gps} and the \lstinline{include_bipm} arguments).

\subsubsection{Handling external data}\label{subsubsec:externaldata}
Performing time and coordinate transformation requires external data such as JPL solar system ephemerides and observatory clock correction files. 
Traditionally, \tp provide copies of these data within the packages, and \tp developers keep them up to date.  However, the upstream data are typically updated frequently, meaning that the \tp developers must often update their packages, and their users must re-install them frequently, rather than simply updating the data directly.
\textsc{Astropy} provides \pint with an easier way to keep these data up to date as many standard timing-related data sets, including but not limited to Earth rotation data, leap seconds, and JPL solar system ephemerides, are updated by \textsc{Astropy}.
For the earth orientation parameters~(i.e., IERS table A and B\footnote{\href{https://datacenter.iers.org/eop.php}{https://datacenter.iers.org/eop.php}}) and solar system ephemerides, \textsc{Astropy} downloads and caches them when requested. 
However, due to the upstream issues, for \textsc{Astropy} versions earlier than 3.2, it requests an upgrade on the package itself to keep the leap seconds up to date, instead of downloading the newest version of leap seconds.
Data not currently handled by \textsc{Astropy}, such as observatory specific clock corrections, are updated by the \pint development team in the traditional manner.  Nonetheless, there are plans for automatic updates of many of these data sets in future \pint releases.

\subsection{Models Module}\label{subsec:model module}
The \pint \lstinline{models} module provides an API for implementing and interacting with pulsar timing models.
In this section, the overall design of the \lstinline{models} module is presented in the beginning, and the public interface object, the \lstinline{TimingModel} class, is discussed after. The details of how to programmaticallly implement a timing model are in Appendix \S\ref{sec:createtimingmodel}. 
Note that this paper does not discuss the implementation of any specific timing model. For these details, please see the online documentation\footnote{\url{https://nanograv-pint.readthedocs.io/en/latest/api/pint.models.timing_model.html}}.

Following the philosophy of modularity, \pint implements different physical effects separately as model components, which are implemented independently in the \lstinline{Component} class and its sub-classes.
Results computed for a timing model are produced by combining the values from the selected components.
The delays produced by each component are simply added together, but for components whose value depends on time --- for example the R\"omer delay depends on the pulsar's position in its orbit --- the time at which each component is evaluated depends on the delays of other components. This requires the components to be computed in a specific order; this order is enforced by PINT but can be overridden by users if necessary (say for custom model components).

A model component implementing a particular mathematical model of a physical effect would be implemented in a sub-class of the base \lstinline{Component} class; this bas class is where the common attributes and functionality of all model components are implemented.
The \lstinline{TimingModel} class is designed to manage the set of included components and provides the overall interface for collecting and returning the results from them, without requiring the calling code to know the details of the specific model.

As described in \S\ref{subsec:models}, modeling TOAs includes two fundamental calculations, total time delay~($\Delta$ in Eqs.~\ref{eq:emission_time} and \ref{eq:total_delay}) and total phase~(Eq.~\ref{eq:phase_number}). 
\pint therefore implements two explicit \lstinline{Component} sub-classes, \lstinline{DelayComponent} and \lstinline{PhaseComponent}.
The \lstinline{TimingModel} class provides two corresponding methods, \lstinline{.delay()} and the \lstinline{.phase()}, to compute the total delay and total phase by adding the results from all the delay and phase components that are included in the model.

\begin{figure}[ht!]
\epsscale{1.3}
\plotone{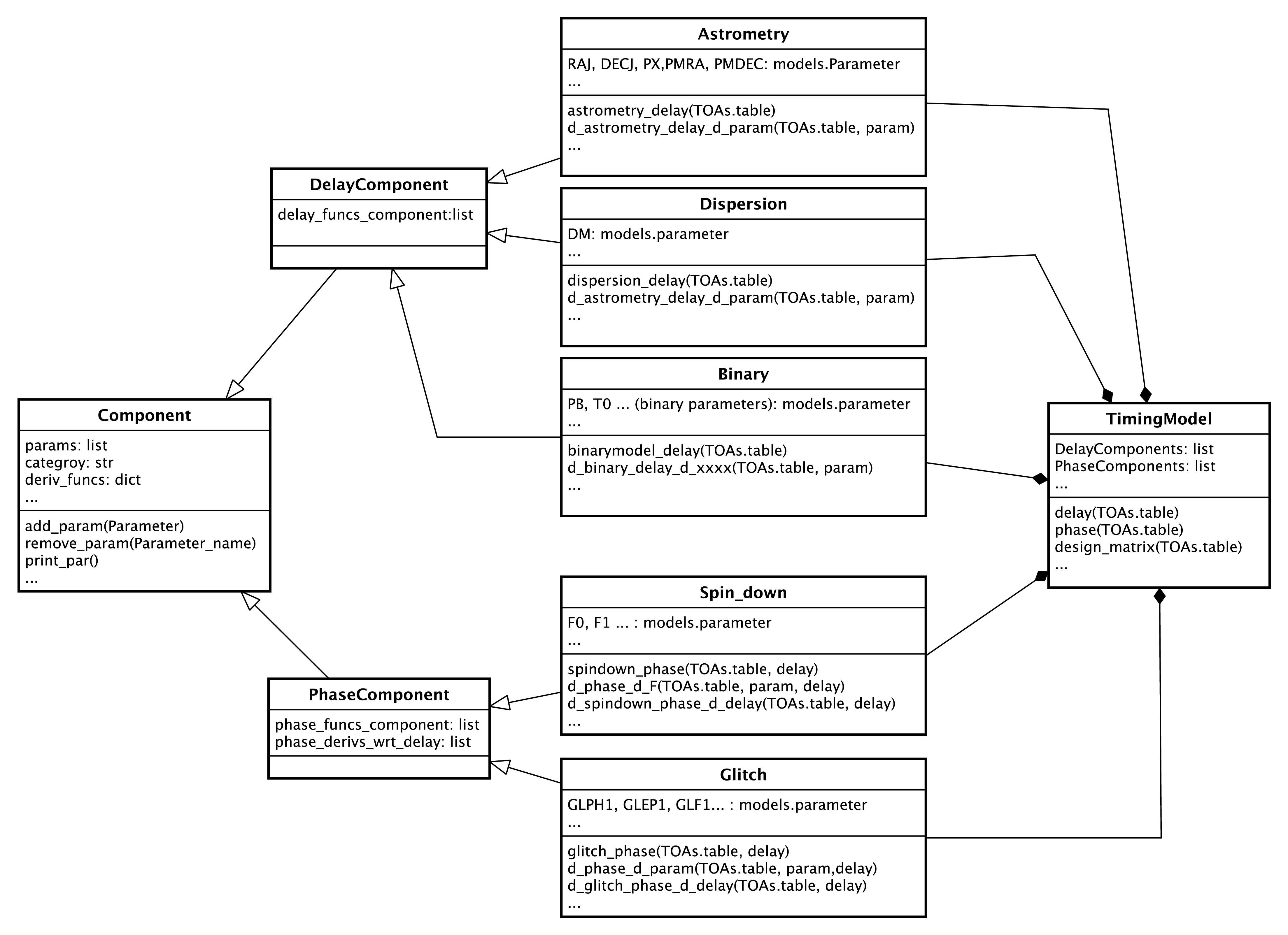}
\caption{This figure shows an example of how \pint implements a full timing model. Hollow arrows indicate ineritance, while solid arrows indicate containment.
\lstinline{Astrometry}, \lstinline{Dispersion} and \lstinline{Binary} classes inherit from the
\lstinline{DelayComponent} class. \lstinline{Spin_down} and \lstinline{Glitch} inherit from
\lstinline{PhaseComponent}. Both \lstinline{DelayComponent} and \lstinline{PhaseComponent} inherit from the generic \lstinline{Component} base class. A \lstinline{TimingModel} instance manages all 
the specific
model components needed to build the full model. 
Here, we only use \lstinline{DelayComponent} and \lstinline{PhaseComponent} as example, if the other component types~(e.g., \lstinline{NoiseComponent}) present, they follow the same relationship structure. \label{fig:timingmodeluml} }
\end{figure}

\pint is not limited to these component types, and is completely extensible to other types. 
For example, \pint also provides a noise model component type, \lstinline{NoiseComponent}, for handling timing noise models used in generalized least squares fitting and Bayesian timing analyses~\citep{timingnoise1,timingnoise2}. 
Similarly, the \lstinline{TimingModel} class also includes the APIs to compute other useful quantities.
For instance, the \lstinline{TimingModel} class is able to compute the design matrix, a key feature needed by the \lstinline{fitter} module, via the \lstinline{.designmatrix()} method.
In Figure~\ref{fig:timingmodeluml}, the layout of the model and component class system is visually illustrated using example model components.

As described in \S\ref{subsec:code architecture}, a \lstinline{TimingModel} object can be initialized via the \lstinline{models.get_model()} function with a \TP-style \textsc{.par} file as input. 
Based on the input \textsc{.par} file, the \lstinline{models.get_model()} function selects and sorts the required components, constructs the \lstinline{TimingModel} object and parses the parameter values. 
More details about the construction of \lstinline{TimingModel} instances are discussed in \S\ref{subsec:model_builder}.
Since version 0.8, \pint{} also provides a wrapper function, \lstinline{models.get_model_and_toas()}, that creates the \lstinline{TimingModel} object and \lstinline{TOA} together from the input \textsc{.par} and \textsc{.tim} files and apply the information in the \textsc{.par} file to \lstinline{TOAs} object. 
Additionally, the \lstinline{TimingModel} object allows users to manipulate the components interactively, beyond simply changing parameter values.  For example, one can change the order of the components or disable individual components.
This design facilitates interactive pulsar timing data processing, which can sometimes be difficult with compiled programs.
A timing model can be adjusted and examined interactively and intermediate computational results can be accessed as needed.

The \lstinline{models} module comes with commonly used timing-model components and functionality. 
Table \ref{tab:built_in_models} lists the built-in model components in \pint\ \version.
For the most updated \lstinline{model} module and built-in components information, please visit our online documentation.

\floattable
\begin{deluxetable}{lllc}
\tablecaption{\pint version \version{} built-in TimingModel categories and components.\label{tab:built_in_models}}
\tablewidth{0pt}
\tablehead{
  \colhead{Model category} & \colhead{Category Description} & \colhead{Component name}  &\colhead{Reference}
}
\startdata
\multirow{2}{*}{\hfil astrometry}& \multirow{2}{*}{\hfil Solar system geometric effects} & AstrometryEquatorial & 1\\
 & & AstrometryEcliptic& 2\\
 \cmidrule{3-4}
solar\_system\_shapiro & Solar system Shapiro delay & SolarSystemShapiro & 3 \\
 \cmidrule{3-4}
\multirow{2}{*}{\hfil dispersion\_model} & \multirow{2}{*}{\hfil Interstellar media dispersion effects}& Dispersion & 4  \\
 & & DMX & 5\\
 \cmidrule{3-4}
 \multirow{5}{*}{\hfil pulsar\_system} & \multirow{5}{*}{\hfil Pulsar system time delay} & BinaryELL1 & 6 \\
 & & BinaryELL1H & 7 \\
 & &BinaryDD & 8 \\
 & &BinaryDDK  & 9 \\
 & &BinaryBT & 10 \\
 \cmidrule{3-4}
 spindown& Spindown phase & Spindown & 11 \\
 glitch& Glitch phase  &  Glitch& 12\\
 frequency\_dependent & Frequency evolution of pulsar profiles & FDdelay & 13\\
 jump& Jump phase offset& JumpPhase & 14\\
 scale\_toa\_error & Template fitting timing noise correction&ScaleToaError& 15\\
 ecorr\_noise&ECORR type noise model&EcorrNoise&16\\
 pl\_red\_noise&Powerlaw red noise type noise model&PLRedNoise&17\\
 ifunc& Interpolated timing noise &IFunc & 18 \\
 wave& Sinusoidal timing noise decomposition & Wave& 19 \\
 solar\_wind & Dispersion due to the solar wind & SolarWindDispersion& 20\\
 troposphere & Delay due to the local atmosphere & TroposphereDelay & 21 \\
\enddata
\tablerefs{(1)(4)(11)~\citet{1986review};
(2)(5)(13)(15)(16)(17)~\citet{NG9yearsdata};
(3)~\citet{shapirodelay};
(6)~\citet{ELL1}; (7)~\citet{ELL1H}; (8)~\citet{DD_model}; (9)~\citet{Kopeikin1995,Kopeikin1996};
(10)~\citet{BT};
(12)(14)~\citet{tempo2};
(18)~\citet{deng12};
(19)~\citet{hobbs10};
(20)~\citet{2006MNRAS.372.1549E};
(21)~\citet{davis1985,niell1996,CRC};}
\end{deluxetable}

\subsection{Residual Module} \label{subsec:residual}
Residuals between the data and timing model are key to updating model parameters and assessing goodness of fit.
The \lstinline{residuals} module is designed to compute the residuals using Eqs.~\ref{eq:phase_res} and~\ref{eq:time_res}.
The interface class, \lstinline{Residuals}, instantiated by providing the \lstinline{TOAs} and \lstinline{TimingModel} instance, implements the \lstinline{.calc_phase_resids()} method and \lstinline{.calc_time_resids()} method for computing the phase residuals and time residuals, respectively. 
For a better representation of the difference between the timing model and the TOAs, the residuals are by default weighted by the TOA uncertainty, but this feature can be switched off in the class method argument.
In addition, if specific pulse numbers are provided, the residuals can be calculated based on those, rather than the nearest integer pulse.
Together with the residual calculation methods, a handful of convenience methods for computing statistics of the residuals are provided (e.g., the $\chi^{2}$ and reduced $\chi^{2}$ values).


\subsection{Fitter Module}\label{subsec:fitter}
The updating of timing models is performed by the \lstinline{pint.fitter} module, which includes a general API base class \lstinline{fitter.Fitter} and a set of pre-defined fitter sub-classes implementing specific optimization algorithms.
The general API base class \lstinline{Fitter} sets up framework, and the fitter sub-classes implement the  fitting algorithms under the  \lstinline{.fit_toas()} class method.
This setup allows the user to implement a new fitting algorithm
with minimum code modifications (only overwriting the \lstinline{.fit_toas()} method), but using the same interface.
Table \ref{tab:fitter_list} lists all the built-in fitters in \pint{} \version{}. Note \pint{}  implements the parameter priors~(see \ref{subsec:parameter}) which is used in the MCMC fitter. The constraints of parameters can be performed via the priors. 
However, the all other fitters do not use this Information other than the MCMC fitter,
and the current fitters can not fit for the noise parameter yet. 
A common package to compute the noise parameter value and parameter priors is \lstinline{enterprise}.

 \floattable
\begin{deluxetable}{cl}
\tablecaption{\pint implemented fitting algorithms
\label{tab:fitter_list}}
\tablewidth{0pt}
\tablehead{
  \colhead{Fitter Name} & \colhead{Algorithm}
}
\startdata
PowellFitter & \textsc{Scipy} Powell minimizing \\
WLSFitter & Weighted least square fitting \\
GLSFitter & Generalized least square fitting \\
MCMCFitter & Markov-Chain Monte Carlo optimization fitting\\
WidebandTOAFitter &  TOAs and independent dispersion measurements joint fitting\tablenotemark{a}
\enddata
\tablenotetext{a}{The independent dispersion measurements are fitted with TOAs simultaneously using generalized least square fitting~\citep{wideband19, wideband20}. }
\end{deluxetable}

As described in the code example in Figure \ref{fig:example_update_model},  a fitter class should be instantiated with \lstinline{TOAs} object and \lstinline{TimingModel} object.
The \lstinline{TimingModel} object will be linked to the \lstinline{fitter.model_init} attribute and an extra copy will be save in the \lstinline{fitter.model} attribute in order to retain initial timing model data.
During fitting, the \lstinline{fitter.model} attribute will be updated but the \lstinline{fitter.model_init} stays the same.
Under this scheme, the original timing model can be easily  traced back by the class method \lstinline{fitter.reset_model()}.
Residuals are calculated and saved in the \lstinline{fitter.resids} attribute, and a copy of initial residuals will be saved to \lstinline{fitter.resids_init} using the same scheme.

One of the most important functionalities of the fitter API is to alter the model parameter information.
The \lstinline{Fitter} base class already provides a set of convenience functions for this purpose.
For example, the \lstinline{.set_params()} class method is designed for changing parameter values and the \lstinline{.set_fitparams()} method can be used for selecting the fitting parameters.

As described above the post-fitting results are returned via the \lstinline{fitter.model} attribute and the \lstinline{fitter.resids} attribute will be updated to post-fit residuals.
This new timing model and residuals are ready for the next iteration.

\section{Comparison of \pint with \TP}\label{sec:compare}

One way to validate \pint is to compare its results with those from the existing high-precision pulsar timing software packages~(i.e., \tp version \tpversion\ and \tpt version \tptversion). In addition to validating \pint, such a comparison checks the accuracy and precision limitations of the various software packages. As of version \version, \pint is capable of analyzing the TOAs from most pulsars, including the \npulsars\ pulsars from the \NG 11-year data release~\citep{NG11yeardata}. Here we present the results of a \pint analysis of PSR~J1600$-$3053 from the \NG 11-year data set, using the DD binary model, including a detailed comparison between \pint and \tp results. PSR~J1600$-$3053 was chosen for this comparison because it has a large number of TOAs (\ntoas) with sub-microsecond timing precision over a long timespan (8 years).  This comparison will also highlight some implementation differences between \pint and \TP. A full scale \pint-\TP comparison using all the pulsars from \NG's 12.5-year data is reported in~\citet{ng12}). The Jupyter notebook for this comparison is included in the \pint examples and can be view from the \pint online documentation\footnote{\href{https://nanograv-pint.readthedocs.io/en/latest/examples-rendered/paper\_validation\_example.html}{https://nanograv-pint.readthedocs.io/en/latest/examples-rendered/paper\_validation\_example.html}}.

\subsection{Comparison using PSR~J1600$-$3053}
We used the published \NG 11-year ephemeris~(originally produced using the \tp software package) as our initial timing model, fitted to TOAs from the \NG 11-year data using the \pint generalized-least-square fitter \lstinline{pint.fitter.GLSfitter}.

\begin{figure}[htb!]
\plotone{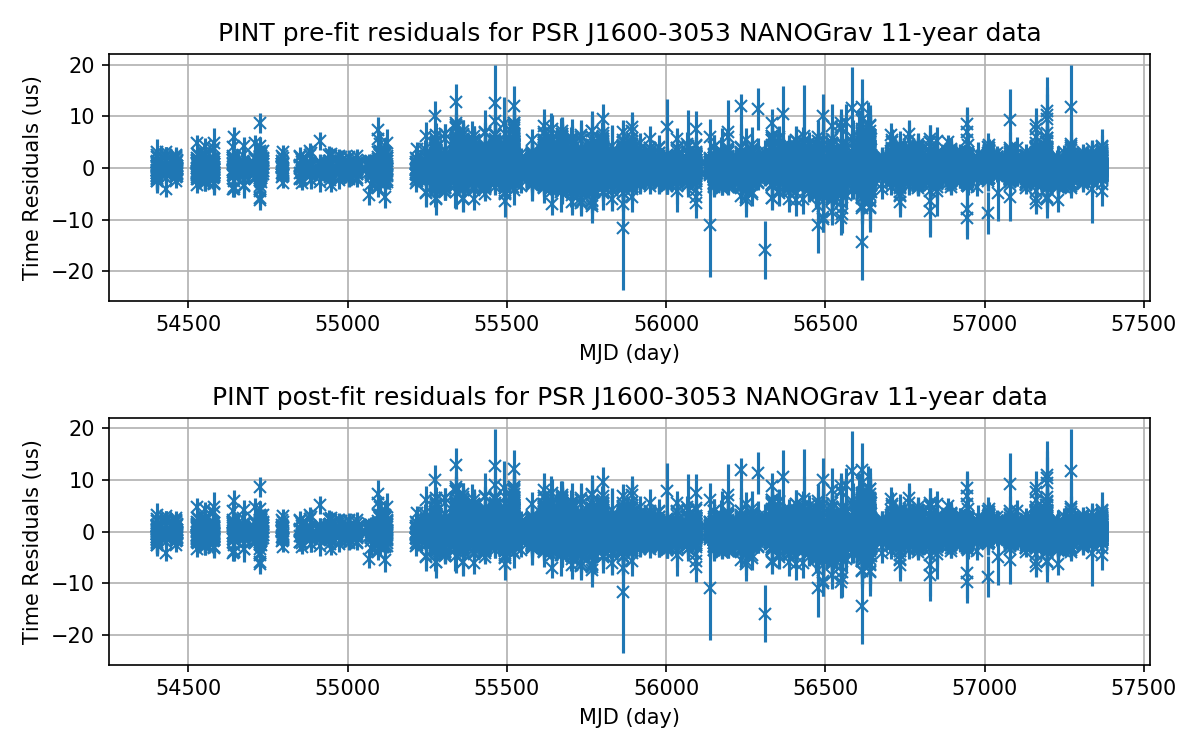}
\caption{Residuals generated by PINT for PSR~J1600$-$3053 from the \NG 11-year data set. The top panel shows residuals before performing a generalized-least-squares fit based on the published \tp-based timing solution. The bottom panel shows the residuals after the fit using \pint.  The RMS of the residuals are nearly identical.}\label{fig:J1600_pint_res}
\end{figure}

The pre-fit residuals from \pint had a weighted-root-mean-square (WRMS) value of \pwrmspre\,$\mu$s. The fitting process reported a final $\chi^{2}$ value of \pchi\ for \dof\ degrees-of-freedom and the post-fit residuals had a WRMS of \pwrmspost\,$\mu$s.
Figure~\ref{fig:J1600_pint_res} shows the \pint pre-fit and post-fit residuals.
In the following subsections, the results of a detailed comparison between \pint and \TP are presented.

\subsubsection{Comparison with \tp results} \label{subsubsec:pint-tp}

\begin{figure}[htb!]
\plotone{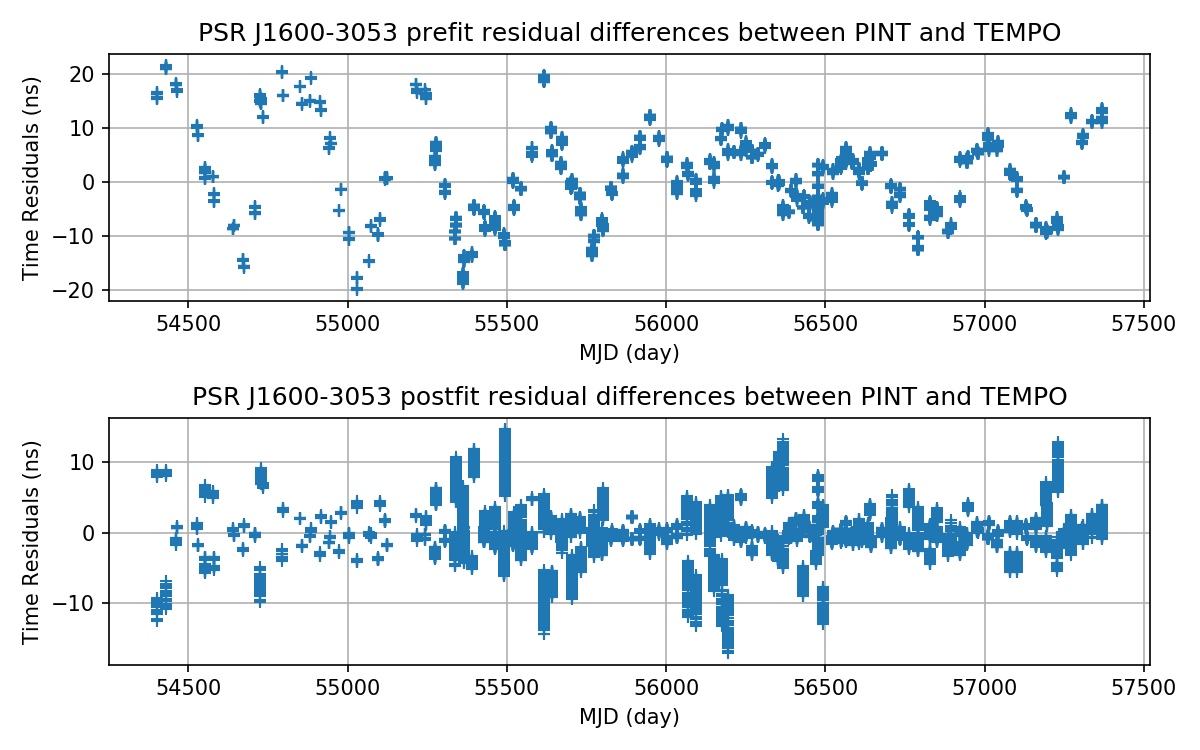}
\caption{Residual differences between \pint and \tp for PSR J1600$-$3053. The upper panel presents the difference of pre-fit residuals and the lower panel presents the post-fit residuals difference. \label{fig:J1600_tempo}}
\end{figure}

The \tp-based fitting for the same data set returns a $\chi^{2}$ value of \tpchi\ and the residuals have a WRMS of \tptwrms\,$\mu$s. We directly compared both the pre-fit and post-fit residuals between these two packages. In Figure \ref{fig:J1600_tempo}, the residual differences between \pint and \tp are presented. Note that since we dropped the constant part of absolute phase in our calculation, a constant offset in the residual differences has been ignored.

\begin{figure}[ht!]
\plotone{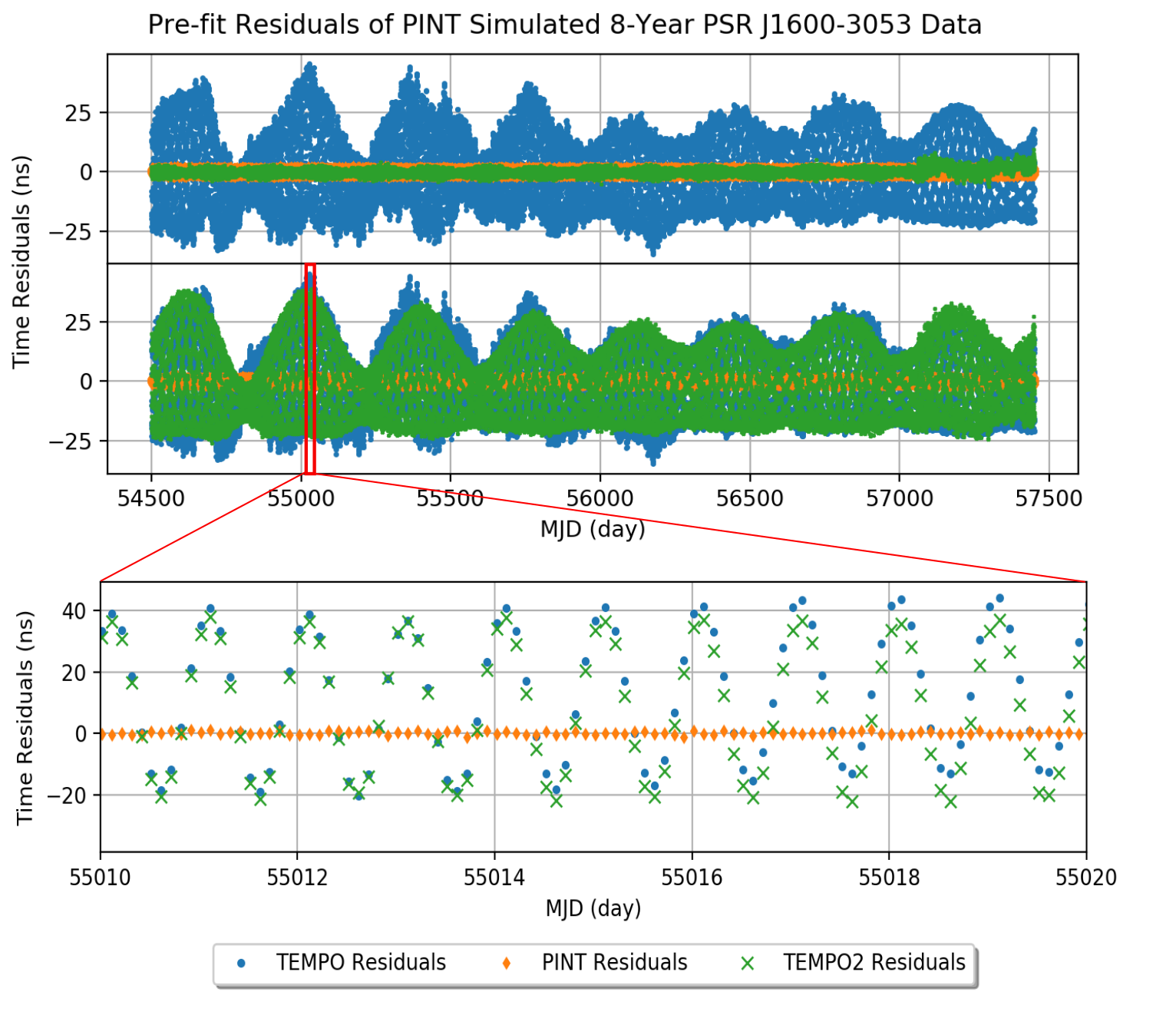}
\caption{The residual difference due to different precession-nutation models. We use \pint to simulate an 8-year regularly sampled (2.4-hour cadence) TOAs with a simple timing model, only has a constant pulse frequency, and pulsar position. The orange marks represent the PINT residuals, the blue points are the \tp residuals, and green data points marks the \tpt residuals. The first panel on the top shows the \pint and \TP residuals when \tpt is under IAU 2000 resolution of precession and nutation. The second panel displays the same results with \tpt's old precession and nutation mode, and \tpt's residuals has a similar signature like \tp residuals. The third panel is a zoomed-in version of the second panel on days from MJD 55010 to MJD 55020. We can see the diurnal sinusoidal oscillation from \TP residuals. Given the sampling rate of \NG 11-year data, the \tp prefit residual differences in Figure \ref{fig:J1600_tempo} is one trace of the blue dots.    \label{fig:simulate_gbt_8y}}
\end{figure}

In the pre-fit residual differences, a distinct annual periodic signature, with a peak amplitude of about 20\,ns is present throughout the whole data set.  This discrepancy is due primarily to different precession-nutation models used in \pint and \tp. \pint uses \textsc{Astropy}'s built-in precession-nutation model~\citep[see the IAU 2000 resolution;][]{IAU2000precission_nutation}, while \tp uses much older models, the IAU 1976 precession~\citep{IAU_1976} and IAU 1980 nutation~\citep{IAU1980}~models. The difference between these models and their impact on timing residuals has been discussed in~\citep{tempo2}. Due to a lack of polar motion in the \tp-style precession-nutation model, the expected timing residual differences should have an amplitude near $\pm$30\,ns with a diurnal signature that is modulated by annual and 435-day periodicities. Figure \ref{fig:simulate_gbt_8y} illustrates the residual discrepancies due to the different precession-nutation models.

We compared the parameters resulting from GLS fits using \tp and \pint as well.  The timing model parameter differences are listed in Table \ref{tab:d_param_pint_tempo}.  All the \pint post-fit parameters are consistent with the \tp parameter values to well within the 1-$\sigma$ uncertainties.  This shows that \pint is capable of reproducing the published result for PSR~J1600$-$3053 in the \NG 11-year data set. 

\floattable
\begin{deluxetable}{lccDDD}
\tablecaption{\pint parameter comparison with \tp for PSR J1600$-$3053
\label{tab:d_param_pint_tempo}}
\tablewidth{0pt}
\tablehead{
  \colhead{Parameter} & \colhead{$V_{\rm T}$\tablenotemark{a}} & \colhead{Unit} & \multicolumn2c{$V_{\rm T} - V_{\rm P}$\tablenotemark{b}}  & \multicolumn2c{$\left| V_{\rm T} - V_{\rm P}\right| /\sigma_{\rm T}\tablenotemark{c}$}  & \multicolumn2c{$\sigma_{\rm P}\tablenotemark{d}/\sigma_{\rm T}$}
}
\decimals
\startdata
F0 & 277.9377112429746(5) &Hz& \se{-1.471}{-14} & 0.028 & 1.000 \\
F1 & \se{-7.33874(5)}{-16} &Hz/second& \se{6.362}{-23} & 0.014 & 1.000 \\
FD1 & \se{4.0(2)}{-5} &second& \se{-2.546}{-9} & 0.002 & 1.000 \\
FD2 & \se{-1.5(1)}{-5} &second& \se{1.370}{-9} & 0.001 & 1.000 \\
JUMP & \se{-8.8(1)}{-6} &second& \se{-4.650}{-10} & 0.004 & 1.004 \\
PX & 0.50(7) &mas& \se{-2.070}{-3} & 0.028 & 1.000 \\
ELONG & 244.347677844(6) &deg& \se{-5.924}{-10} & 0.099 & 1.000 \\
ELAT & $-$10.07183903(3) &deg& \se{-3.191}{-9} & 0.095 & 1.000 \\
PMELONG & 0.46(1) &mas/year& \se{7.119}{-4} & 0.068 & 1.003 \\
PMELAT & $-$7.16(6) &mas/year& \se{-5.048}{-4} & 0.009 & 0.999 \\
PB & 14.348466(2) &day& \se{-3.457}{-8} & 0.016 & 1.000 \\
A1 & 8.8016531(8) &light-second& \se{1.491}{-8} & 0.018 & 0.984 \\
A1DOT & \se{-4.0(6)}{-15} &light-second/second& \se{8.913}{-18} & 0.014 & 1.000 \\
ECC & \se{1.73729(9)}{-4} &dimensionless& \se{-2.386}{-10} & 0.027 & 1.002 \\
T0 & 55878.2619(5) &day& \se{-1.051}{-5} & 0.020 & 0.991 \\
OM & 181.85(1) &deg& \se{-2.638}{-4} & 0.020 & 0.991 \\
OMDOT & \se{5(1)}{-3} &deg/year& \se{-2.211}{-5} & 0.016 & 1.000 \\
M2 & 0.27(9) &Solar Mass& \se{-1.641}{-3} & 0.018 & 0.979 \\
SINI & 0.91(3) &dimension less& \se{5.436}{-4} & 0.016 & 0.984 \\
$\rm{DMX\_0010}$\tablenotemark{e} & \se{6(2)}{-4} &$\rm{pc/cm}^{3}$& \se{-5.089}{-6} & 0.025 & 1.000 \\
\enddata
\tablenotetext{a}{\tp fit parameter value.}
\tablenotetext{b}{\pint fit parameter value.}
\tablenotetext{c}{\tp fit parameter uncertainty.}
\tablenotetext{d}{\pint fit parameter uncertainty.}
\tablenotetext{e}{In the \NG 11-year data, PSR J1600$-$3053 has 106 DMX time ranges. Here we only list the one DMX parameter that has the largest difference between \pint and \tp.}
\end{deluxetable}

\subsubsection{Comparison with \tpt results} \label{subsubsec:compare_tp2}

Prior to the \pint-\tpt comparison, we modified the timing model parameter files from the published \NG 11-year data set for a more controlled comparison. The 11-year data set timing models used \tp, which has adopted the ecliptic coordinate frames with the 2010 IAU value of the obliquity ~\citep{NG9yearsdata}. However, \tpt implements the ecliptic coordinate frame using the 2003 IAU obliquity value. Thus, we chose to use the 2003 IAU obliquity value in this comparison. Another modification is due to the discrepancy in the precession and nutation model mentioned in the previous section. Fortunately, \tpt allows for the user to choose between the IAU 2000 resolution and the \tp style precession-nutation model~\citep{tempo2}. Naturally, we decided to run \tpt under the same precession-nutation model (IAU 2000 resolution) as \pint.

\begin{figure}[htb!]
\plotone{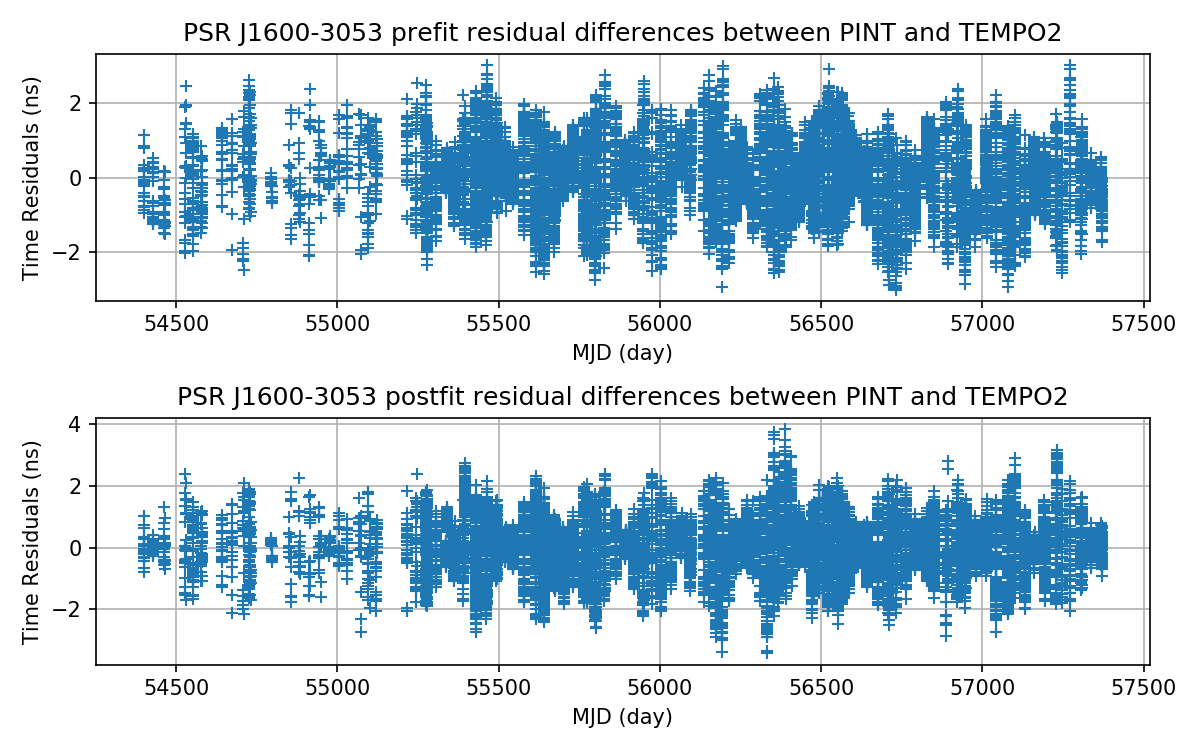}
\caption{Residual difference between \pint and \tpt for the J1600$-$3053 \NG 11-year data. The upper panel shows the pre-fit residual difference and the lower panel shows the post-fit residual difference.\label{fig:PINT_TEMPO2}}
\end{figure}

\begin{figure}[ht!]
\plotone{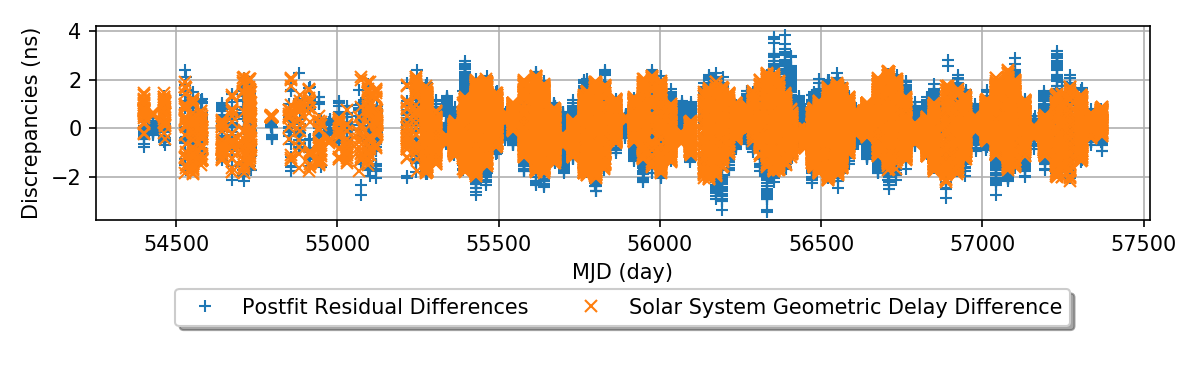}
\caption{\pint-\tpt residual differences and the \pint-\tpt solar system geometry delay difference plotted on top of each other. The blue data points mark the difference between \pint and \tpt post-fit residuals and the orange points mark the difference between \pint and \tpt solar system geometric delay. Their envelopes trace with each other, show that the 2\,ns level residual discrepancies are caused by the solar system geometric delay implementation difference of these two softwares.   \label{fig:roemer_diff}}
\end{figure}

\tpt's generalized-least-squares fitting gives a final $\chi^{2}$ value of \tptchi\ and the post-fit residuals have a WRMS of \tptwrms\,$\mu$s. \tpt residuals were also directly compared against the \pint residuals, and the comparison is shown in the Figure \ref{fig:PINT_TEMPO2}. Again a constant residual offset has been ignored here as well.
Both the pre-fit and post-fit residual differences are less than 10\,ns, which is within the accuracy goal of \tpt~\citep{tempo2}. However, the residual differences show systematic quasi-periodic signature with a semi-annual term that occurs consistently over the whole data set. The same signature presents in the \pint-\tpt solar system geometric delay (i.e.,~R{\o}mer Delay) difference as well. In Figure \ref{fig:roemer_diff}, the solar system geometric delay difference and the residual differences are plotted together. This common signature indicates that the 2.5\,ns level residual discrepancies are due to a difference in the solar system geometric delay calculation~(e.g., observatory position or pulsar sky location). We also compared the post-fit parameters between \pint and \tpt, and all agree within the \tpt-reported parameter uncertainties (see Table \ref{tab:d_param_pint_tempo2}).

\floattable
\begin{deluxetable}{lccDDD}
\tablecaption{\pint fit parameter vs the \tpt parameter
\label{tab:d_param_pint_tempo2}}
\tablewidth{0pt}
\tablehead{
  \colhead{Parameter} & \colhead{$V_{\rm T2}$\tablenotemark{a}} & \colhead{Unit} & \multicolumn2c{$V_{\rm T2} - V_{\rm P}$\tablenotemark{b}}  & \multicolumn2c{$\left| V_{\rm T2} - V_{\rm P}\right| /\sigma_{\rm T2}$\tablenotemark{c}}  & \multicolumn2c{$\sigma_{\rm P}\tablenotemark{d}/\sigma_{\rm T2}$} 
  }
\decimals
\startdata
F0 & 277.9377112429746(5) &Hz& \se{-6.661}{-16} & 0.001 & 1.000 \\
F1 & \se{-7.33874(5)}{-16} &Hz/second& \se{-8.192}{-24} & 0.002 & 1.000 \\
FD1 & \se{4.0(2)}{-5} &second& \se{-1.636}{-9} & 0.001 & 1.000 \\
FD2 & \se{-1.5(1)}{-5} &second& \se{1.416}{-9} & 0.001 & 1.000 \\
JUMP & \se{-8.7887456483}{-6} &second& \se{-4.904}{-11} & 0.0004\tablenotemark{f} & N/A \\
PX & 0.50(7) &mas& \se{1.878}{-5} & 0.0003 & 1.000 \\
ELONG & 244.347677843(6) &deg& \se{9.123}{-12} & 0.002 & 1.000 \\
ELAT & $-$10.07183905(3) &deg& \se{-1.449}{-11} & 0.0004 & 1.000 \\
PMELONG & 0.46(1) &mas/year& \se{7.420}{-6} & 0.0007 & 1.000 \\
PMELAT & $-$7.16(6) &mas/year& \se{-7.171}{-5} & 0.001 & 1.000 \\
PB & 14.348466(2) &day& \se{-1.924}{-09} & 0.0009 & 1.000 \\
A1 & 8.8016531(8) &light-second& \se{-8.197}{-10} & 0.001 & 1.000 \\
A1DOT & \se{-4.0(6)}{-15} &light-second/second& \se{-1.034}{-18} & 0.002 & 1.000 \\
ECC & \se{1.73730(9)}{-4} &dimensionless& \se{4.159}{-11} & 0.005 & 1.000 \\
T0 & 55878.2619(5) &day& \se{4.883}{-7} & 0.0009 & 1.000 \\
OM & 181.84(1) &deg& \se{1.226}{-5} & 0.0009 & 1.000 \\
OMDOT & 0.005(1) &deg/year& \se{-1.229}{-6} & 0.0009 & 1.000 \\
M2 & 0.27(9) &Solar Mass& \se{1.043}{-4} & 0.001 & 1.000 \\
SINI & 0.91(3) &dimensionless& \se{-4.278}{-5} & 0.001 & 1.000 \\
$\rm{DMX\_0099}$\tablenotemark{e} & 0.0017(2) &$\rm{pc/cm}^{3}$& \se{-3.773}{-7} & 0.002 & 1.000 \\
\enddata
\tablenotetext{a}{\tpt post-fit parameter value.}
\tablenotetext{b}{\pint post-fit parameter value.}
\tablenotetext{c}{\tpt post-fit parameter uncertainty.}
\tablenotetext{d}{\pint post-fit parameter uncertainty.}
\tablenotetext{e}{In the \NG 11-year data, PSR J1600$-$3053 has 106 DMX time ranges. Here we only list the DMX parameter with the largest discrepancy between two packages.}
\tablenotemark{f}{Since this version of \tpt did not report the JUMP uncertainty. The relative difference is computed using the \pint fit uncertainty, and the uncertainty division is not applicable.}
\end{deluxetable}

\subsection{Other known implementation differences between \pint and \TP}
In this section we present four major known implementation differences between \pint and \tp that could cause substantial differences in the results. We show differences in the timing between \pint and \tp for several other pulsars presented in the \NG 11-year data set.
\begin{description}
\item[UTC(GPS) to standard UTC clock conversion (\tp only)] As described in \S\ref{subsec:pintcoords}, \pint converts UTC(GPS) time to the standard UTC timescale.
However, the \tp package does not apply this 10-nanosecond-level clock correction to the TOAs.
In Figure \ref{fig:gps correction}, the UTC(GPS) and standard UTC clock correction values over the past two decades are plotted.
\begin{figure}[htb!]
\plotone{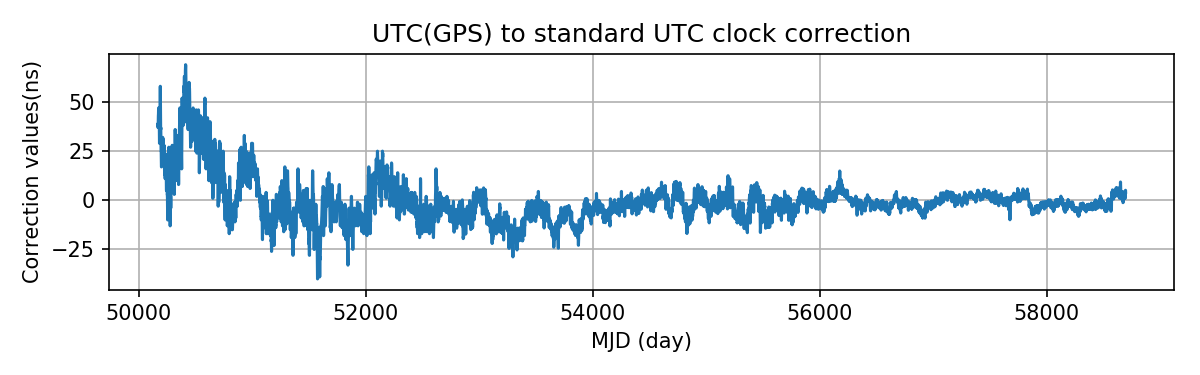}
\caption{UTC(GPS) and standard UTC clock correction over 20 years since the GPS timescale was established. \label{fig:gps correction}}
\end{figure}

\item[Constant time offset between TOAs correction (\texttt{JUMP})]
The constant time offset between TOAs, implemented as \texttt{JUMP}s in the timing model, can be introduced from two major effects:
(1) a constant time delay from different instruments~(e.g., different cable length), and (2) pulse-profile evolution delays~(e.g., from the frequency evolution of the intrinsic pulse profile).
Since the first type of time offset occurs at the observatory, it should be corrected at the observatory frame~(before computing the solar system barycentric TOAs). 
The pulse profile offset is a part of the intrinsic pulsar emission process.
Thus, the second type of \texttt{JUMP}s is more appropriately applied under the pulsar frame.
However, both \tp and \tpt do not distinguish these two type of \texttt{JUMP}s and correct both of them under the same reference frame. 
\tp corrects the \texttt{JUMP}s in either observatory frame or the pulsar frame~(\tp gives the options to the user). 
\tpt applies the \texttt{JUMP} corrections in the pulsar frame in terms of phase offset. 
In this release of \pint, the \texttt{JUMP}s are applied in the same way as the \tpt method. 
However, \pint has infrastructure to apply the two types of \texttt{JUMP}s separately, and it is planed in the future releases. 
Therefore, if \tp corrects the \texttt{JUMP}s at the observatory, a highly radio-frequency-dependent residual discrepancy with a period of one year will be present in the \pint-\tp residuals difference (see Figure \ref{fig:jump_diff}).
The peak value of this yearly signature is dependent on the \texttt{JUMP} offset values.

\begin{figure}[htb!]
\plotone{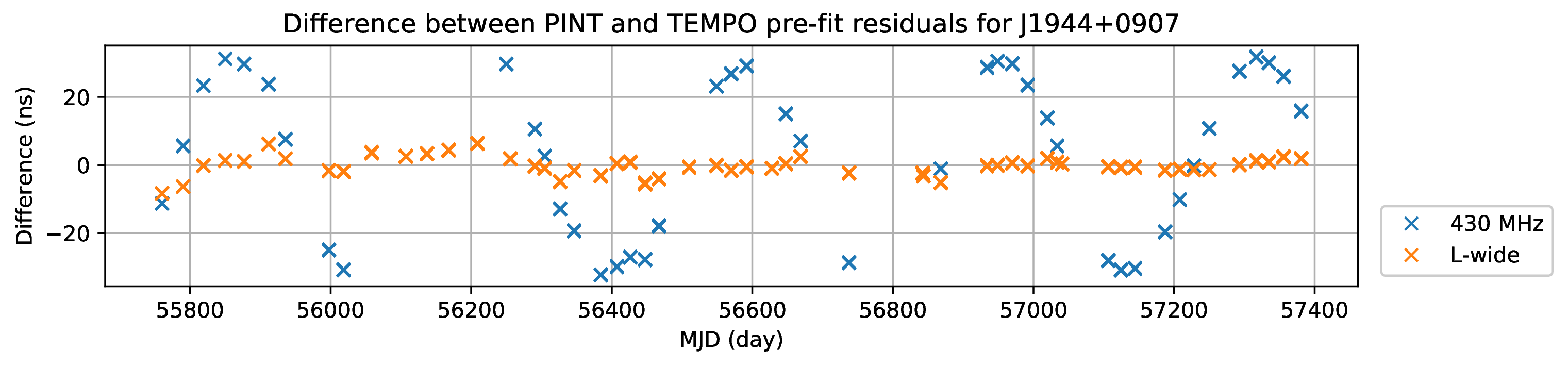}
\caption{Residual difference between \pint and \tp pre-fit residuals for PSR J1944+0907 \NG 11-year data. This discrepancy introduced by different \texttt{JUMP} calculations. 
Since, the \texttt{JUMP}s in \tp are applied on the 430\,MHz receiver, the annual sinusoid variations only show up for the 430\,MHz TOAs.\label{fig:jump_diff}}
\end{figure}

\item[Frequency-dependent delay~(FD delay)]
The frequency-dependent delay is implemented for modeling the pulse profiles variation at different radio frequencies by \NG ~\citep{NG9yearsdata}.
Instead of applying the FD delay before the pulsar binary correction like \TP, \pint applies it to the TOAs after the binary model in the pulsar frame.
This delay introduces an offset in the binary model input TOAs, which leads to a $\sim 10$\,ns level of residual difference, which depends on the FD parameter values (see Figure \ref{fig:fd_diff}).
\begin{figure}[htb!]
\plotone{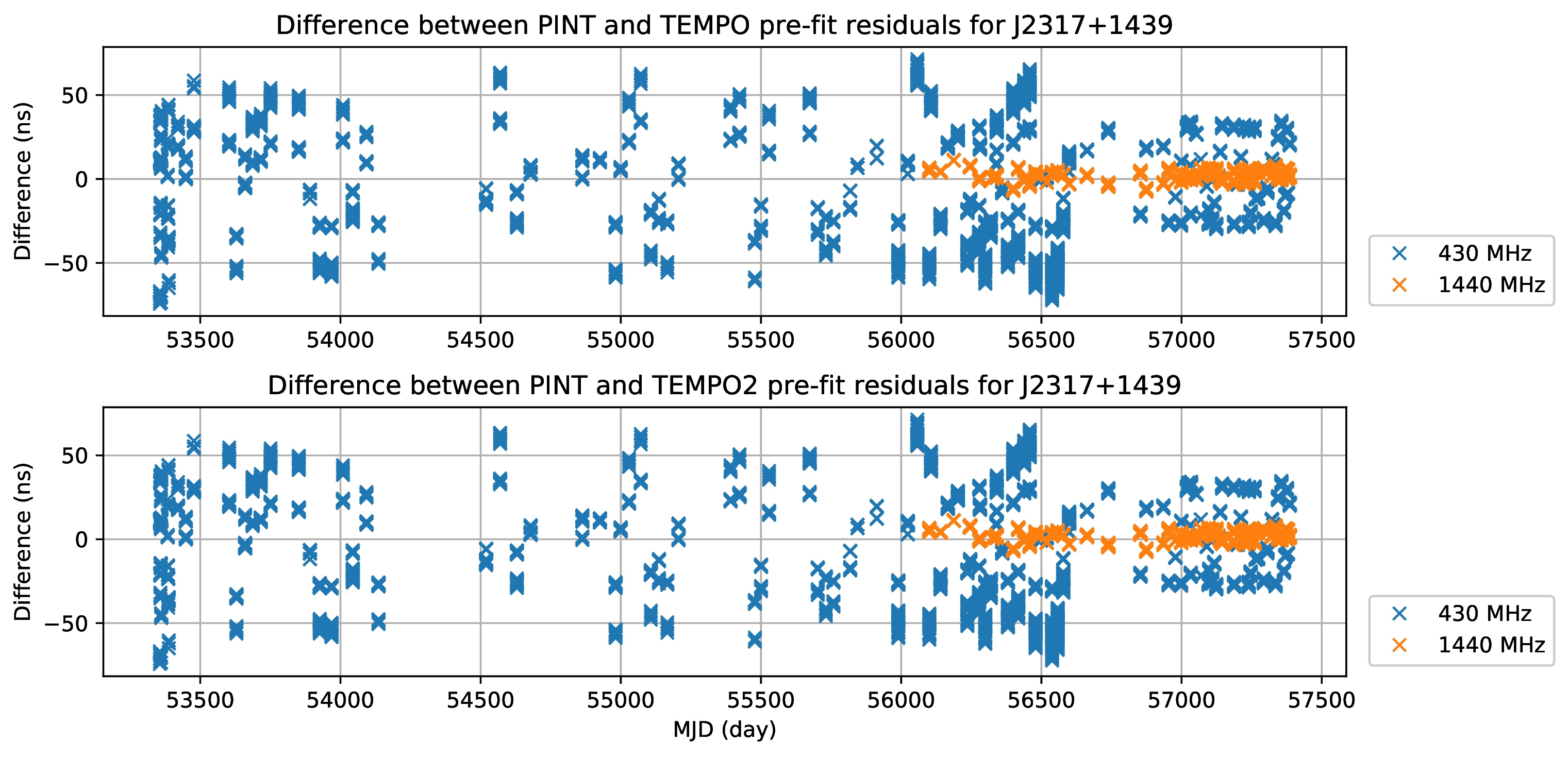}
\caption{Residual differences between \pint and \tp due to a discrepancy in the radio Frequency-Dependent delay~(FD delay). The first panel illustrates the PSR J2317+1439 \NG 11-year data \pint-\tp residual difference, and the second panel illustrates the \pint-\tpt residual differences for the same data set. The radio frequency band, 1440\,MHz, residual differences are marked in orange, and the band of 430\,MHz residual differences are marked in blue. The 430\,MHz shows a higher variation on the difference plot. Because the FD delay is higher at the lower frequency band, this leads to a bigger discrepancies in the binary delay input TOAs. Since \tp and \tpt both apply the FD delay before the binary correction, these two results are very similar, so that both panels show almost identical plots. \label{fig:fd_diff}}
\end{figure}
\end{description}

Aside from the difference mentioned above, \pint uses a uniform definition of the longitude of ascending node, known as the ``\texttt{KOM}'' parameter in DDK binary model ~\citep{Kopeikin1995,Kopeikin1996}, which is measured with respect to equatorial north.
In \TP, the \texttt{KOM} parameter is defined with respect to the north of the reference frame under which the pulsar position is given~(i.e., if the pulsar position is given as ecliptic coordinate, \texttt{KOM} parameter is measured from ecliptic north).

\subsection{Independence from \TP}\label{subsec:independence}
One of the motivations of the  \pint project is to provide independent (or as independent as is reasonably possible) cross checks and/or validation of the timing results from other pulsar timing packages. For high-impact precision timing programs, such as gravitational wave detection efforts, it is critical to compare results from more than a single data analysis pipeline.

\pint is not a Python wrapper of other code, nor is it a Python translation of C or FORTRAN code from previous timing packages.
The framework, APIs, and internal data storage are implemented independently.
The fundamental algorithms, such as linear algebra,  solar system coordinate transformations, and unit conversions, are from widely used and well-tested public python packages (e.g., \textsc{NumPy}, \textsc{Astropy}).
\pint's built-in models are implemented based on the physical formulas from their respective publications, and the detailed references are incorporated in the code documentation~(e.g., the equation numbers from the papers and necessary derivations are documented in the documentation strings and/or source code).
This re-implementation automatically provides a cross-check to the same models as implemented in, for example, \TP.  
When validating the built-in models, we compare \pint's results~(e.g., residual and post-fit parameter values and uncertainties, or direct calculations of delay times, for example) with \TP, and attempt to resolve all the discrepancies by auditing both packages' code and their references carefully. 
This is how we identified implementation differences described in \S \ref{sec:compare}, as well as long-standing bugs in \tpt related to planetary Shapiro delays\footnote{See \url{https://bitbucket.org/psrsoft/tempo2/issues/63/incorrect-planetary-shapiro-delays}.} and the solar angle calculation\footnote{See \url{https://bitbucket.org/psrsoft/tempo2/issues/68/sign-error-in-solar-angle-calculation}.}.
Aside from comparing the same physical model with different implementations, \pint's flexibility, such as being able to call model components from the Python command line, enables the user to easily test or compare algorithms and implementations with other versions in \pint or with other software.

Despite these differences in implementation, \pint adopts most current standard pulsar timing conventions, including data formats and the usage of external data~(e.g., the JPL solar system ephemerides and standard clock correction files).
\pint supports most \TP-accepted styles of TOA and parameter files, and attempts to provide as much backwards compatibility as is reasonably possible. 
This allows users to cross-check or reproduce earlier results without changing their input data formats. 
There are plans to include additional compatibility options in future releases of \pint, such as timing using the INPOP solar system emphemerides series\footnote{\href{https://www.imcce.fr/recherche/equipes/asd/inpop/}{https://www.imcce.fr/recherche/equipes/asd/inpop/}}~\citep{inpop} or with reference to TCB rather than TDB time.

\section{Performance, testing, and maintenance}\label{sec:unittest_mainten}
The \pint project's goal is to provide a high precision, reliable, relatively efficient (i.e.,~fast), and user friendly software package.
To achieve this goal, we require a comprehensive test suite, profiling, effective version control and other development practices, and good documentation.
In this section, we discuss the \pint's performance, testing and maintenance in detail.

\subsection{Performance}\label{subsec:performance}

Compared to compiled languages, one potential drawback of using a high-level interpreted language like Python is execution speed. In particular, there is a substantial startup cost for a Python script as all the necessary packages are imported, and portions of the code that do a lot of looping and object creation are slower than for compiled languages. However, \pint{} makes use of highly optimized vectorized code from \textsc{NumPy} and \textsc{SciPy} for array and linear algebra operations, and can save 
intermediate results, such as the TOAs table as a Python ``pickle'' file, which can be loaded
very quickly. Thus the relative performance depends on the particular problem and how \pint{} is used. 
In this sub-section, we report the \pint run-time for a typical use case of loading a model and TOAs and fitting and compare it with that of \tp and \tpt.
We chose two test cases: (1) a simple timing model for PSR NGC6440E, which includes astrometry, dispersion, and spindown components, comprising 5 free parameters, and (2) a more complex timing model for PSR J1910+1256 from the \NG 12.5-year data set, with 13 model components and 103 free parameters. These were run on Intel(R) Core(TM) i7-10510U CPU @ 1.80GHz, Ubuntu 20.04.1 LTS VM w/ 8GB RAM. Different computers and software libraries will give different results. To see the script used to generate these tables, please visit our GitHub page.\footnote{See \url{https://github.com/nanograv/PINT/tree/master/profiling}}

Table \ref{tab:simple_performace_compare} lists the run-time of \pint and \TP for the case of PSR NGC6440E~(with the same timing model and the same fitter as the code example in Fig. \ref{fig:example_update_model}) with different numbers of simulated TOAs.  
Given the efficiency of FORTRAN and C/C++, \tp and \tpt are faster and more RAM efficient than \pint for small problems dominated by reading TOAs from text files and doing preprocessing (applying clock corrections and computing positions and velocities of the observatory and solar system bodies).
The \pint \lstinline{TOAs} object's pickling functionality allows users to read in TOAs and process them once, save the results to a binary file, and then perform multiple fits or other operations. 
Table ~\ref{tab:timing_breakdown} shows the breakdown of the \pint run-time for different parts of the problem.
Reading from TOA object pickle files is 2--30 times faster than parsing the TOA text files.
 
\floattable

\begin{deluxetable}{lrrrr}
\tablecaption{Performance comparison between \pint, \tp, and \tpt for a simple model\tablenotemark{a}
\label{tab:simple_performace_compare}}
\tablewidth{0pt}
\tablehead{
  \colhead{} & \colhead{\tp} & \colhead{\tp2} & \colhead{\pint}  & \colhead{\pint} \\
  \colhead{Number of TOAs} &\colhead{(second)} & \colhead{(second)} &\colhead{No Pickling} & \colhead{Using Pickle} \\
  &&&(second)&(second)
}
\startdata
100 & 0.250 & 1.194 & 2.174 & 1.894 \\
1,000 & 0.288 & 1.320 & 3.346 & 1.954 \\
10,000 & 0.426 & 1.680 & 17.020 & 3.054 \\
100,000 & 1.972 & 6.370 & 151.170 & 12.734\\
\enddata
\tablenotetext{a}{Averaged over five runs.}
\end{deluxetable}

\floattable
\begin{deluxetable}{lrrrr}
\tablecaption{\pint timing breakdown\tablenotemark{a,b}
\label{tab:timing_breakdown}}
\tablewidth{0pt}
\tablehead{
    \colhead{} & \colhead{Import} & \colhead{Loading TOAs} &  \colhead{Loading TOAs} & \colhead{Fitting} \\
    \colhead{Number of TOAs} & \colhead{Statements} & \colhead{No Pickling} & \colhead{Using Pickle} & \colhead{WLSFitter} \\
    \colhead{} & \colhead{(second)} & \colhead{(second)} & \colhead{(second)} & \colhead{(second)}
}
\startdata
100  & 1.476 & 0.471 & 0.010 & 0.120 \\
1,000 & 1.476 & 2.098 & 0.096 & 0.143 \\
10,000 & 1.476 & 14.961 & 1.037 & 0.432 \\
100,000 & 1.476 & 162.165 & 12.332 & 2.818 \\
\enddata
\tablenotetext{a}{These times were recorded separately from the runs in Table ~\ref{tab:simple_performace_compare}, and there are additional, smaller operations not displayed. Thus, there may be small disparities in timing between the summation of these individual parts and the total runtime recorded in Table ~\ref{tab:simple_performace_compare}.}
\tablenotetext{b}{Averaged over five runs.}
\end{deluxetable}

For the case of PSR J1910+1256 with the complicated timing model, we use the \NG 12.5-year data set's TOAs~(5012 TOAs in total) and timing parameters~(103 free parameters). 
We fit data using generalized least square (GLS) fitting with noise parameters. 
To test the speed of a large number of TOAs within the modeled time span, we duplicated the TOAs 2 times and 5 times.
As seen in Table ~\ref{tab:complex_performance_compare}, the GLS fitting in \TP, coupled with a more complex model, can increase runtime significantly. When using the GLS fitter, execution time will depend  on the linear algebra libraries (i.e., LAPACK) installed and the configuration of the respective software packages. 
In the case of large numbers of TOAs, \pint{} GLS fitting outperforms \TP{}. This could be due to  different linear algebra libraries, or  different implementations of GLS fitting algorithm in these packages.


\floattable
\begin{deluxetable}{lrrrr}
\tablecaption{Complex model for PSR J1910+1256 performance comparison between \pint, \tp, and \tpt\tablenotemark{a,b}
\label{tab:complex_performance_compare}}
\tablewidth{0pt}
\tablehead{
  \colhead{} & \colhead{\tp} & \colhead{\tpt} & \colhead{\pint} & \colhead{\pint} \\
  \colhead{Number of TOAs} & \colhead{(second)} & \colhead{(second)} & \colhead{No Pickling} & \colhead{Using Pickling} \\
  \colhead{} & \colhead{} & \colhead{} & \colhead{(second)} & \colhead{(second)}
}
\startdata
5,012 & 32.644 & 24.630 & 42.636 & 35.972 \\
10,024 & 249.492 & 52.394 & 60.458 & 47.206 \\
25,060 & 3695.400 & 211.972 & 119.190 & 79.730 \\
\enddata
\tablenotetext{a}{GLS Fitter is used for above runs.}
\tablenotetext{b}{Averaged over five runs.}
\end{deluxetable}

To aid current and future optimization efforts, PINT comes with a folder of profiling code, allowing users and developers to see both a general summary and a detailed report of how long it takes PINT to perform tasks. These files make use of cProfile, Python’s built-in profiling tool. Users and developers can produce flow charts to visualize where PINT spent the most time and find bottlenecks in the code. An html viewer (independent of PINT and cProfile) for the cProfile output is also available, allowing the user to click into a function and see the subsequent functions called. Thus, the user can find the root function consuming the most time, or a function taking an unexpectedly long time, and optimize the embedded code. It is our hope that with these features, PINT will become faster as more and more people use the profiling features. The authors themselves have been able to reduce certain benchmark speeds by over 15\% using these features.

\subsection{Testing}\label{subsec:unittest}
\pint{} provides various scripts for testing the package, most of which are systematically executed before incorporating any change into the code base.
The aim of this testing is to ensure reliability and reproducibility, but \pint code that is never run as part of the test suite is certainly not being checked.
As version of \version, \testcover\ of the code is executed during these tests, and increasing this fraction, as well as ensuring that tests check essential properties, is a goal for future releases.
For any development and modification, running the test scripts helps detect potential bugs that may break other \pint{} modules, or, ideally, user code.
Thus, providing testing code for new features is strongly encouraged.  
In order to maintain the package's stability and compatibility,  the \pint project has adopted the on-line and off-line testing tools,
\lstinline{pytest}\footnote{\href{https://docs.pytest.org/en/latest/}{https://docs.pytest.org/en/latest/}}, \lstinline{hypothesis}\footnote{\href{https://github.com/HypothesisWorks/hypothesis}{https://github.com/HypothesisWorks/hypothesis}},
\lstinline{GitHub Actions}\footnote{\href{https://github.com/features/actions}{https://github.com/features/actions}}, and \lstinline{tox}\footnote{\href{https://tox.readthedocs.io/en/latest/}{https://tox.readthedocs.io/en/latest/}}. These tools execute our tests on the major UNIX based operating systems with different \lstinline{Python} versions. 

\subsection{\pint maintenance}\label{subsec:PINTmaintain}
Following the design philosophy of ``for and by the user'', the \pint software package is an open source project under the BSD 3-clause license\footnote{\href{https://opensource.org/licenses/BSD-3-Clause}{https://opensource.org/licenses/BSD-3-Clause}}.
A user can develop and modify \pint software freely as long as the copyrights are recognized.

Since \pint is an ongoing development project, it adopts a modern version control scheme using \lstinline{git} and GitHub\footnote{\href{https://git-scm.com/}{https://git-scm.com/}, \href{https://github.com/}{https://github.com/}}.
The GitHub page~(\url{https://github.com/nanograv/PINT}) is where the \pint software official versions are released and where a user can communicate with the development team, open issues and propose changes through pull requests. The \pint user manual can be found at the link above as well. 
We encourage the user community to contribute to the \pint project by submitting pull requests and reporting issues.

The documentation is compiled in Restructured Text format using standalone text files and the document strings inside the python code, using \lstinline{Sphinx}\footnote{\url{http://www.sphinx-doc.org}}. 
Each time a change is merged into the \lstinline{master} branch, the documentation is deployed to \url{readthedocs.io} where it is automatically compiled and made available as a website (\url{https://nanograv-pint.readthedocs.io}).

\section{Example \pint use cases}\label{sec:use_cases}

Fundamentally, \pint is a \lstinline{Python} library that users can employ to do pulsar timing calculations in \lstinline{Python} scripts
or Jupyter\footnote{\url{http://jupyter.org}} notebooks of their own creation. As such, \pint is now included as a dependency in other \lstinline{Python} timing libraries (e.g. 
\NG's \lstinline{enterprise}\footnote{\url{https://github.com/nanograv/enterprise}}; 
\lstinline{stingray}\footnote{\citet{huppenkothenStingrayModernPython2019}; \url{github.com/stingraysoftware/stingray}}; 
\lstinline{HENDRICS}\footnote{\citet{bachetti_hendrics:_2018}; \url{github.com/stingraysoftware/HENDRICS}}).

However, several common use cases have been implemented as command-line \lstinline{Python} scripts that are distributed with \pint, serving as examples and allowing many users to employ \pint\
without needing to explicitly write \lstinline{Python} code:
\begin{description}
\item[\texttt{pintempo}] A command-line script that provides similar functionality to the \tp and \tpt programs. It reads a timing model and TOAs from specified files and fits parameters, optionally making a residuals plot.

\item[\texttt{pintbary}] A simple script for barycentering (i.e. converting to TDB timescale and applying Solar System delays) specified times, allowing specification of the observatory and observation frequency.

\item[\texttt{pintk}] A graphical user interface inspired by the \texttt{plk} plugin for \tpt. Users can modify the model and TOAs, perform fits, revert to previous fits, and view the results on residuals plot with a choice of axes. The interface is highly interactive and subsets of TOAs can be selected for fitting. In addition, JUMPs and phase wraps can be easily added and removed without changing the parfile or timfile. As an aid for phase connection, \texttt{pintk} can also plot sets of random models with parameters drawn from  the covariance matrix of each fit to see how well a model extrapolates across data gaps.

\item[\texttt{zima}] A script to generate a set of simulated TOAs based on an input timing model.
\end{description}
In addition to these applications, there are also scripts included that are specific to handling high-energy
(X-ray, $\gamma$-ray) photon data, as described below.

\subsection{High Energy Photon Timing}

\pint has a number of tools that enable processing of photon data by treating the arrival time of each photon event as a TOA. These are often from space-borne X-ray and $\gamma$-ray telescopes.
The biggest difference between these events and traditional TOAs is that they are not expected to have occurred at a fiducial phase; they have some distribution in phase, and the goal of the project may even be to determine whether there is any evidence of phase dependency in this distribution. More, these events are often taken from an observatory that is in orbit
and thus not at a fixed ITRF coordinate like a ground-based observatory. \pint's \lstinline{observatory} module smoothly handles
these cases, as described in section \S\ref{subsubsec:observatory}.  
\pint is able to handle events from \textsc{FITS} files that contain
unmodified spacecraft times, or those that have been barycentered or geocentered by mission-specific software such as
\lstinline{gtbary}~\citep{fermitool} or \lstinline{barycorr}~\citep{nicertool}. For unmodified spacecraft times, the
relevant \lstinline{Observatory} class is initialized with a (mission-specific) orbit file that contains data on the position of the spacecraft
as a function of time. \pint builds a univariate spline interpolator that allows for easy computation of the spacecraft
position (and velocity) at the precise time of any photon event.  Given this, the rest of the \pint machinery can be used
on these data.  Such data sets often contain large numbers of events, so this often puts a premium on efficient, vectorized computations, made possible by the \textsc{NumPy} arrays that \pint uses.

Here again, these functions are available for use as \lstinline{Python} modules, but several common use cases have been
implemented as command-line scripts distributed with \pint:
\begin{description}
\item[\texttt{photonphase}] A code that reads common X-ray event data (e.g., from \textit{NICER}, \textit{XMM/Newton}, \textit{NuSTAR},  \textit{RXTE}) from
\textsc{FITS} files and computes the pulse phase of each event using a provided timing model.  The output can be plotted or written back
out to a column in a \textsc{FITS} file.
\item[\texttt{fermiphase}] A code similar to \lstinline{photonphase} that is specific to \textit{Fermi} $\gamma$-ray data. One addition
is the ability to handle photon weights.
\item[\texttt{event\_optimize}] A code that demonstrates fitting a pulsar timing model to photon data, using \pint to compute model
phases and \texttt{emcee}\citep{emcee} to perform an MCMC maximum-likelihood optimization.
\end{description}

The \textit{NuSTAR} team is using PINT for the new clock correction pipeline (Bachetti et al. in prep.).
Recently, the Very-High-Energy(VHE) $\gamma$-ray community has been investigating the use of \pint as part of their
processing pipelines. Their data are photon events from ground-based observatories.

\section{Conclusion and discussion}
High-precision pulsar timing experiments, including ground-based and space-based projects, are now monitoring a large number of pulsars regularly~(for example, \NG monitored \npulsars\ millisecond pulsars for its 11-year data release).
Around the globe, thousands of precisely measured TOAs are generated using high sensitivity radio telescopes and their modern receivers and backends~(wideband receivers and GPU-based backends, etc.) every year.
These efforts aim to detect new, extreme astrophysical signals, like the low-frequency stochastic gravitational-wave background.
However, it has been very challenging to analyze these large and intricate data sets and share them between international pulsar timing groups~(see e.g., \citealt{IPTADR1}, as each group uses their own tools to record and analyze data).
In addition, historical data sets are still very valuable for current and future timing projects~(e.g., comparing the differences between instruments).
This requires that an analysis pipeline has sufficient backwards compatibility.  

We present the \pint software package, which provides a platform to overcome these challenges by using an object-oriented and modular design, adopting well-debugged \lstinline{Python} libraries, and incorporating the modern version control tools \lstinline{git} and GitHub. The \pint package is capable of processing high-precision pulsar timing data with a numerical precision of $\sim$1\,ns and with algorithmic precision of a few ns or better.

We briefly summarize the code architecture and four core modules \lstinline{toa}, \lstinline{models}, \lstinline{fitter}, and \lstinline{residuals} module.  
\begin{itemize}
 \item \lstinline{toa} module provides the functionality of storing and pre-processing~(i.e., applying clock corrections and computing the observatory location and velocity) the TOAs from different observatories.
 \item \lstinline{models} module maintains a set of built-in model components and the public interface class, \lstinline{TimingModel}, for interacting and organizing the model components.
The model component class, \lstinline{Component}, and its sub-classes provide the infrastructure for implementing a new model with minimum effort and for performing pulsar data analysis smoothly.

\item \lstinline{fitter} module provides the infrastructures for fitting a model to a set of TOAs and allows implementing a new fitting algorithm routine without modifying the main code.

\item \lstinline{residuals} module implements the container class, \lstinline{Residuals} class, for storing timing residuals and their statistical attributes and methods.   
\end{itemize}

A comparison between \pint and \TP packages is presented in this paper. After the general-least-square fitting on the same test data set, \pint's post-fit parameters are consistent with the results from \TP, within their \TP fit uncertainties, and \pint post-fit residuals differ from \tp and \tpt result at the level of 10\,ns and 1\,ns, respectively.
Some known sources of the discrepancies are described.  

We also demonstrate the unique features of \pint. 
\pint modules and functions are designed as an interactive data analysis platform where the user has access to each step of internal calculation.
Since \pint is a \lstinline{Python}-based package, importing other packages provided by the \lstinline{Python} community becomes extremely simple.  
This innovation creates the possibility for applications or features that are hard to implement with the traditional software packages.
Using the modern version control tool \lstinline{git} and the powerful online interface of GitHub, \pint developers are able to communicate with \pint users and provide technical support.
Along with the package, some convenient command-line scripts are also provided for the common use cases. 
In future releases, the \pint project will keep providing new features and improvements of the code.

\acknowledgments
This project was initiated and supported by the \NG{} collaboration, which receives support from NSF Physics Frontiers Center award number 1430284. The National Radio Astronomy Observatory is a facility of the National Science Foundation operated
under cooperative agreement by Associated Universities, Inc. Portions of this work performed at NRL were supported by Office of Naval Research 6.1 funding. Student research at NRL was sponsored by the Office of Naval Research NREIP program.
SMR is a CIFAR Fellow.
RvH was supported by NASA Einstein Fellowship grant PF3-140116.

%



\software{\texttt{Astropy} ~\citep{2013A&A...558A..33A},~\texttt{emcee}, \texttt{\tp}, \texttt{\tpt} ~\citep{tempo2}, \texttt{git}, \texttt{NumPy}}

\bibliographystyle{aasjournal}
\bibliography{PINT}

\appendix
\section{Create a timing model Component}\label{sec:createtimingmodel}
\begin{figure}[ht!]
\begin{mdframed}
\includegraphics[scale=0.6]{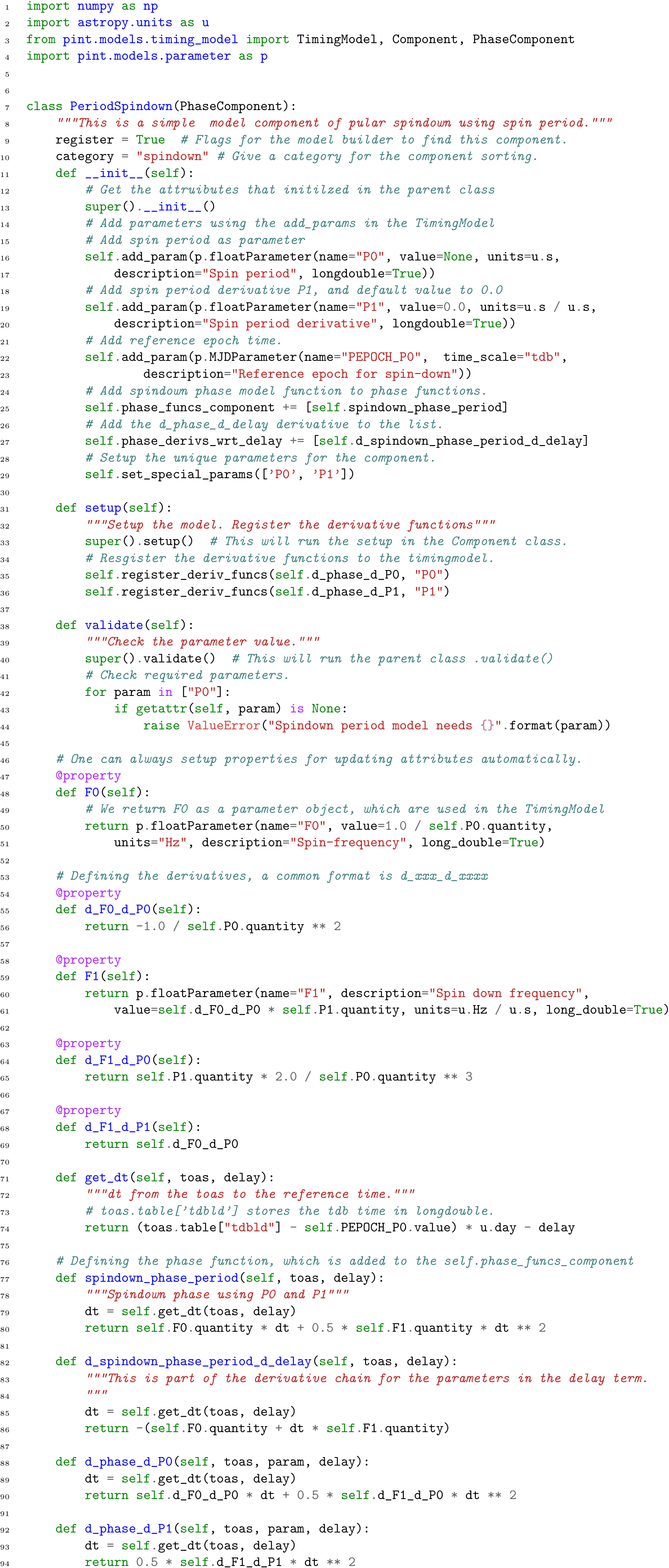}
\end{mdframed}
\caption{Example implementation of a timing model component for pulsar spin-down. \label{fig:simple_spindown} }
\end{figure}

\pint\ is designed to be expandable to new models and new features. 
We encourage our users to build custom models for their needs. 
Here, we present the ingredients of a new timing model component.
The mechanics of automatic model building are in this section as well.
A brief code example is provided in Figure~\ref{fig:simple_spindown} to illustrate how to implement a complete \pint model component that can interact with the \lstinline{TimingModel} class. along with the descriptions. 
A detailed example for composing a model component is included in our online documentation\footnote{\href{https://nanograv-pint.readthedocs.io/en/latest/examples/How\_to\_build\_a\_timing\_model\_component.html}{https://nanograv-pint.readthedocs.io/en/latest/examples/How\_to\_build\_a\_timing\_model\_component.html}}.

A typical timing model component includes three major parts, model parameters (see \S\ref{subsec:parameter} for more details), model functions, and derivative functions.
Model parameters, implemented by the \lstinline{Parameter} class, represent the astrophysical quantities the model depends on (e.g., the pulsar sky locations (\texttt{RAJ}, \texttt{DECJ}), the dispersion measure (\texttt{DM}) and the pulsar pulse frequency (\texttt{F0}), etc.).
The model functions are where model output quantities (e.g., delay, phase, or noise effects) are computed.
The derivatives of modeled quantities with respect to the parameters are required for many fitting algorithms, and so the derivative functions are provided to compute these.

To allow the \lstinline{TimingModel}'s high-level methods to collect the result from the model component, two API conventions must be followed: 1) the returned result has to be in the accepted format, and 2) the model function must to be registered.
For instance, \lstinline{DelayComponent} must return delays as an \lstinline{astropy.units.quantity} object with time units.  
This allows \lstinline{TimingModel.delay()} to sum all the delays correctly without explicit unit conventions needing to be followed in the code.
For \lstinline{PhaseComponent}, the final result should be a \lstinline{pint.phase.Phase} object, which represents pulse phase at the required precision.
In addition, the model functions must be added to the appropriate function lists.
The \lstinline{TimingModel} computes the modeled quantity by sequentially summing the results of the functions in these lists.
Taking the same example, the delay/phase model functions should be added to \lstinline{.delay_funcs_component} or \lstinline{.phase_funcs_component} lists in the \lstinline{delayComponent} or \lstinline{phaseComponent} classes, respectively.

The model component class is also responsible for providing derivative functions with respect to the parameters.
To enable the \lstinline{TimingModel} class to compute the derivatives using high-level wrapper functions, \lstinline{d_delay_d_param()} and \lstinline{d_phase_d_param()} for example, \pint implements a registration scheme for derivative functions.
This scheme requires all derivative functions follow a consistent API; that is, these functions should have specific input arguments and return values (e.g., the phase derivatives should have the TOA table, parameter name, and total delay as the input arguments).
When setting up a model component, derivative functions should be registered using the \lstinline{Component.register_deriv_funcs()} class method which maps the parameter to its derivatives.
The \lstinline{TimingModel} class computes the derivatives by enumerating the derivative functions with respect to the target parameter from all the model components, and then summing the result from these derivative functions.
Users are encouraged to provide accurate derivative functions; fitters that depend on these derivatives may fail completely or converge very slowly if they are wrong or inaccurate. Other fitters, like those based on Markov Chain Monte Carlo algorithms, my not use the derivatives at all but often run much more slowly. However, if analytic derivatives are not provided, approximate derivatives can be obtained automatically by numerical methods in \lstinline{TimingModel.d_delay_d_param_num()}  or \lstinline{TimingModel.d_phase_d_param_num()} with appropriate differential steps.
In the case of phase derivatives, the \lstinline{d_phase_d_param()} also applies the derivative chain rule~(i.e., the phase is first differentiated with respect to delay, and then times the delay derivative with respect to the parameter).
If applicable, the phase derivative with respect to delays should be provided in the phase component.

\subsection{Parameter module}\label{subsec:parameter}

Information about the parameters of a timing model is stored in instances of the \lstinline{Parameter} class and its sub-classes defined in the \lstinline{models.parameter} sub-module.  
These collect all information relevant to a specific model parameter, including its value, uncertainty, units and description (see Table~\ref{tab:param_attr} for a list of key attributes). There is a profusion of subclasses of Parameter in order to handle the variety of different types and formats that parameters can have (for example, strings, right ascensions, floating-point), and also to handle extensible families of parameters like the pulse frequency derivatives F0, F1, \ldots, or like JUMP parameters which select subsets of the arrival time measurements to apply time delays to.

One of the innovative features of the \lstinline{Parameter} class is programmatic integration between a parameter's value and its units.
The \lstinline{.quantity} attribute saves the parameter value as an \lstinline{astropy.unit.Quantity} object, or compatible type of object (e.g., \lstinline{astropy.time.Time}), which contains the physical units and allows automatic unit conversions when performing arithmetic with other quantities.
This feature avoids confusion and errors arising from unit conversions having to be manually implemented in the code. 
Each parameter's uncertainty is saved in the \lstinline{.uncertainty} attribute using the same scheme.
For calculations that do not require unit information, the raw numerical parameter and uncertainty values can still be accessed via the \lstinline{.value} and \lstinline{.uncertainty_value} properties; these are always guaranteed to return the numerical value in the units specified in the \lstinline{.units} attribute.
The parameter value and uncertainty can be changed by setting the \lstinline{.quantity} and \lstinline{.uncertainty} attribute, with unit conversions handled automatically, or \lstinline{.value} and \lstinline{.uncertainty_value}. 

To read a parameter's information from a \textsc{.par}-style parameter file, the \lstinline{Parameter} class provides the \lstinline{.from_parfile_line()}  method, which parses the parameter file line that has the matching parameter name.
The \lstinline{Parameter} class also implements the \lstinline{.as_parfile_line()} method to write a parameter as a \textsc{.par}-style string line.

Another advanced feature is that the parameter's prior probability density function can be set at the \lstinline{.prior} attribute for Bayesian timing parameter estimation (e.g., Markov chain Monte Carlo(MCMC) fitting~\citealt{mcmcfitting}).

\floattable
\begin{deluxetable}{lll}
\tablecaption{Parameter class key attributes \label{tab:param_attr} }
\tablewidth{0pt}
\tablehead{
  \colhead{Attribute} & \colhead{Description}}
\startdata
\lstinline{name} & Parameter name \\
\lstinline{aliases} & Aliases (alternative names) for the parameter  \\
\lstinline{units} & Default unit of the parameter \\
\lstinline{description} & Description of the parameter\\
\lstinline{quantity} & Parameter quantity (with units)  \\
\lstinline{value} &  Parameter numerical value in the default unit  \\
\lstinline{prior} & Prior probability distribution for the parameter \\
\lstinline{uncertainty} & Post-fit parameter uncertainty (with units) \\
\lstinline{uncertainty\_value} & Parameter uncertainty numerical value in the default unit\\
\lstinline{frozen} & Boolean flag for turning on/off fitting of the Parameter
\enddata
\end{deluxetable}

\begin{figure}[ht!]
\begin{mdframed}
\plotone{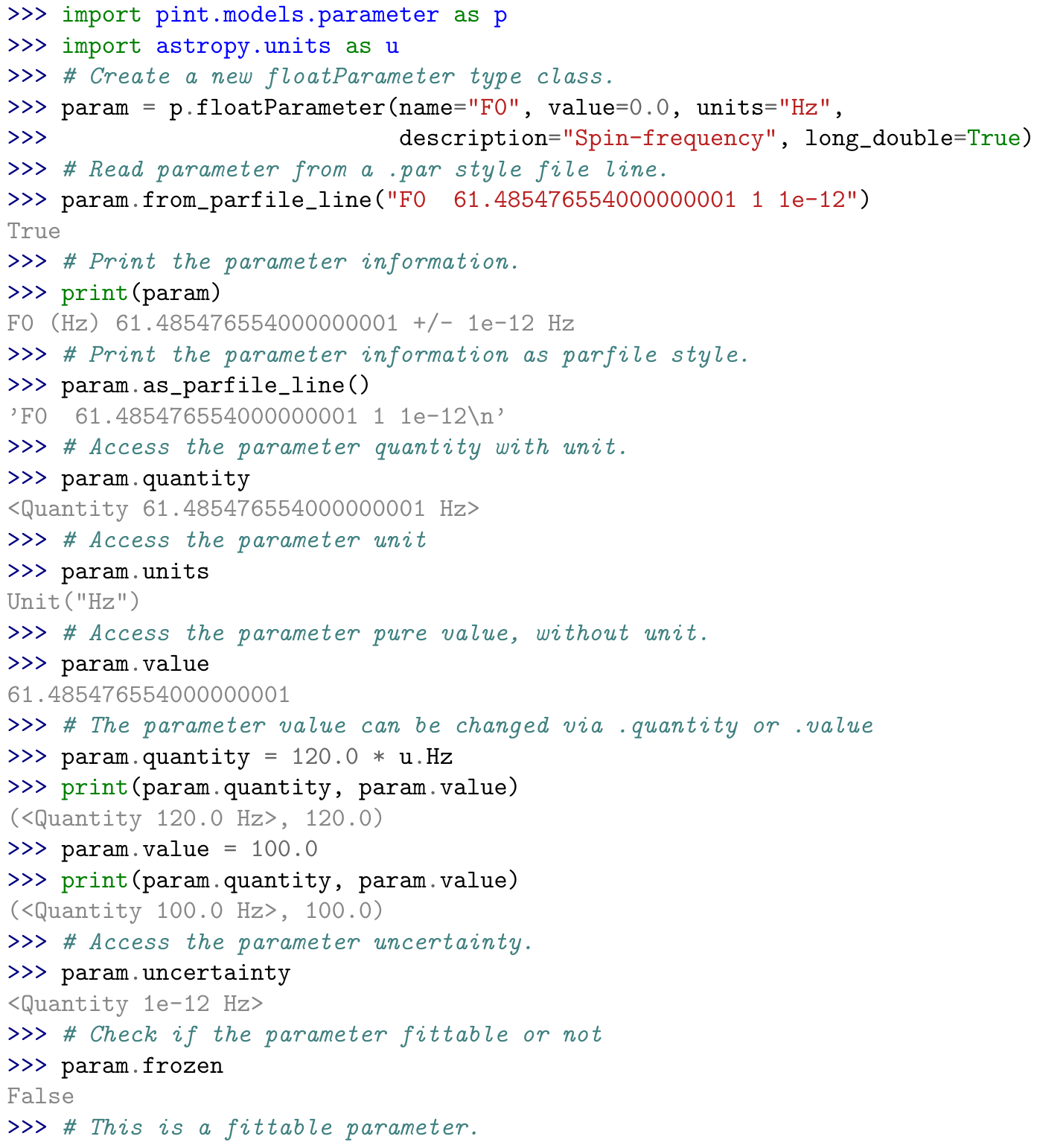}
\end{mdframed}
\caption{Code example for parameter module\label{fig:code_example_parameter}}
\end{figure}

In pulsar timing analysis, timing model parameters are applied to more use cases than typical numerical parameters. 
For instance, the ``BINARY'' parameter represents the binary model name as a string. 
Thus, in \pint, a set of \lstinline{Parameter} sub-classes for different use cases are also implemented.
In the section below, the parameter types provided in this release are listed.
\begin{description}
\item[\texttt{floatParameter}] A parameter type for storing floating-point values. 
The data are stored as an \lstinline{astropy.units.quantity} object, and the precision can be either the 64 bit float or \lstinline{np.longdouble}.
\item[\texttt{strParameter}] A parameter object to store a string value.
\item[\texttt{boolParameter}] A type of parameter object used as Boolean flags. It is able to recognize different format of Boolean value (e.g., `Y/N', `YES/NO' or `1/0')
\item[\texttt{MJDParameter}] A parameter type created for the Modified Julian Day time values. In order to keep the precision and allow a convenient timescale transformation, it is stored as the \lstinline{astropy.time.Time} object.
\item[\texttt{AngleParameter}] A parameter type implemented for the astronomical angle parameters (e.g., Right Ascension or Declination). The parameter value is saved in the \lstinline{astropy.coordinates.Angle} object which provides angle conversion functions. This object accepts different input angle format as well (e.g., `hour:minute:second' or `degree:minute:second')
\item[\texttt{PrefixParameter}] A parameter type designed for parameters that have the same name prefix but a different suffix. (e.g., ``DMX\_0001'' and ``DMX\_0002''). Since this object is implemented according to the parameter name not the value type, it is able to store any other \lstinline{Parameter} types (e.g., MJDParameter, AngleParameter). 
These internal parameter types can be specified via its input argument \lstinline{parameter_type}.
\item[\texttt{maskParameter}] This parameter object provides functionality for parameters that apply only to a subset of TOAs (e.g. a JUMP). It accepts different parameter values like the \lstinline{PrefixParameter} object as well. It is able to handle a parameter that has a key value pair for selecting TOAs (e.g., ``\lstinline{ECORR -f Rcvr1_2_GASP   0.00370}'', for example applying an ECORR value only to TOAs with a particular flag).
\end{description}
Although the \lstinline{Parameter} objects introduced above can be initialized and used independently (see the code example in Figure~\ref{fig:code_example_parameter}), it is recommended to use the \lstinline{Component.add_param()} class method
to add the \lstinline{Parameter} object into the \lstinline{Component} object and register it to the parameter name space.
This allows the automatic model builder (discussed below in \S\ref{subsec:model_builder}) to select model components by comparing the parameter names.

 \begin{figure}[ht!]
\begin{mdframed}
\plotone{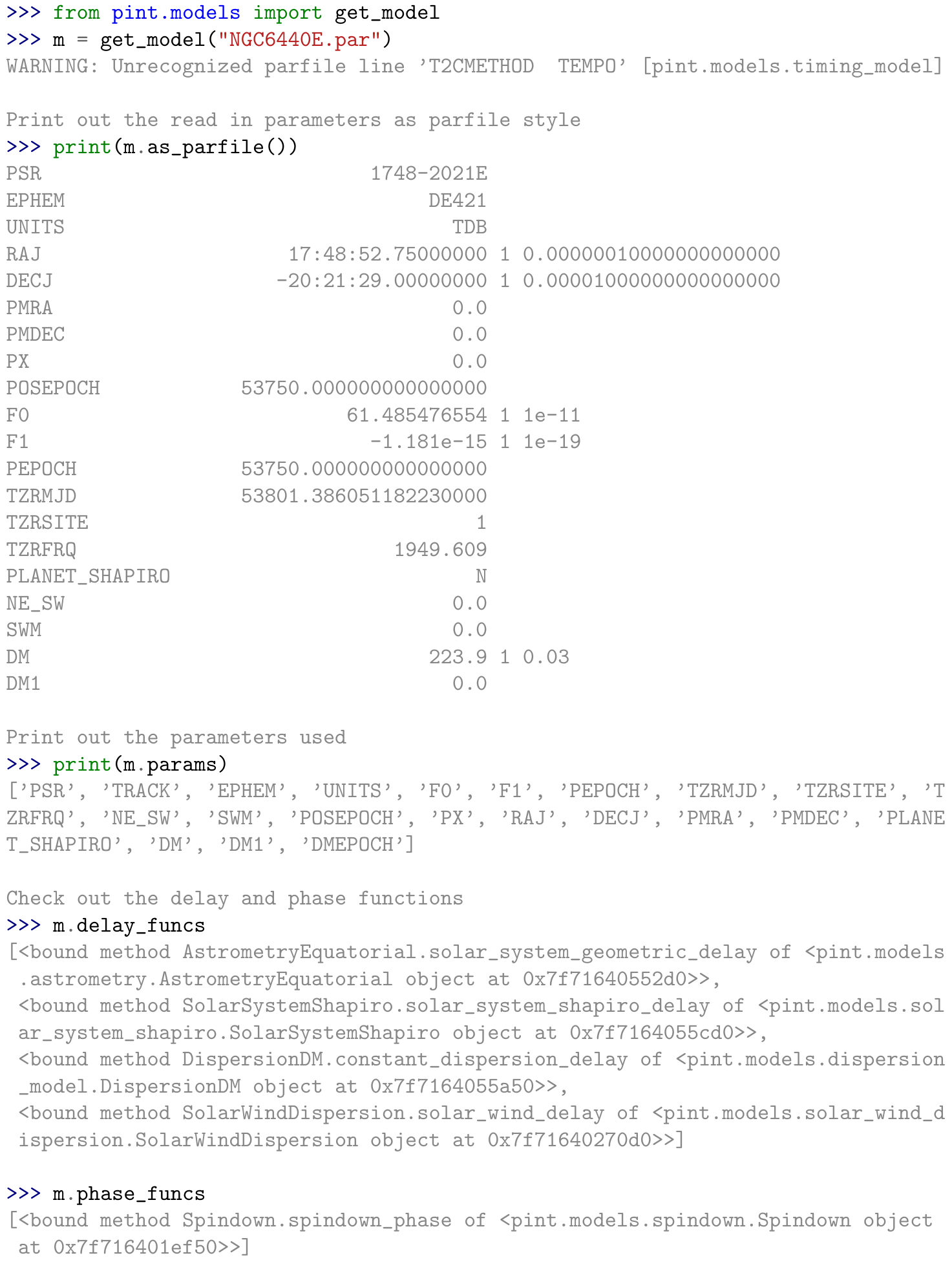}
\end{mdframed}
\caption{Code example for Timing Model module}
\label{fig:TimingModel_code}
\end{figure}

\subsection{Connecting components to the \lstinline{TimingModel}}\label{subsec:model_builder}
In order to properly instantiate the various timing model components, including for example, properly registering the partial derivative functions used by \pint\ for fitting, a user will typically use the  \lstinline{get_model()} function (introduced in \S\ref{subsec:models}), which utilizes the \lstinline{model_builder} module and associated \lstinline{ModelBuilder} class behind the scenes.
The \lstinline{model_builder} selects the correct model components and sorts them into a preferred order, and reads the input parameter values.
The \lstinline{model_builder} searches for all registered model components, whose attribute \lstinline{.register} is set to be True, as demonstrated in the code example in Figure~\ref{fig:simple_spindown} (see line 10).
After listing all the components, it compares each component's parameters with the parameters in the \textsc{.par} file,
and When they are in common, the component is selected.
However, this method has two challenges that could lead to a wrong model selection: (1) One astrophysical effect can be modeled using different parametrization~(e.g., the DM variation can be modeled by a Taylor expansion or a set of discrete DM values). (2) Different components may share a set of common parameters (e.g., some more complicated components are derived from a simple components).
To help the \lstinline{model_builder} filter the components, \pint implements a component category system and a special parameter identifier.
\lstinline{model_builder} reads the component's category from the component attribute \lstinline{.category}, and only one component from the same category will be selected.
For instance, even \pint{} has five built-in model components in the $pulsar\_system$ category, but one timing model can only make use of a pulsar binary component.  
As of \pint \version, we classify all the components in the categories listed in Table~\ref{tab:built_in_models}.
Each model component specifies its unique parameters in the \lstinline{.component_special_params} attribute and the \lstinline{model_builder} will first check if these unique parameters are specified in the \textsc{.par} file.
In the end, the selected components are sorted by category and model parameter values are read in.



\end{document}


\section*{}

\begin{Verbatim}[commandchars=\\\{\},numbers=left,firstnumber=1,stepnumber=1]
\PY{k+kn}{import} \PY{n+nn}{numpy} \PY{k}{as} \PY{n+nn}{np}
\PY{c+c1}{\PYZsh{} PINT uses astropy units in the internal cacluation.}
\PY{k+kn}{import} \PY{n+nn}{astropy}\PY{n+nn}{.}\PY{n+nn}{units} \PY{k}{as} \PY{n+nn}{u}

\PY{c+c1}{\PYZsh{} Import the component classes.}
\PY{k+kn}{from} \PY{n+nn}{pint}\PY{n+nn}{.}\PY{n+nn}{models}\PY{n+nn}{.}\PY{n+nn}{timing\PYZus{}model} \PY{k+kn}{import} \PY{n}{TimingModel}\PY{p}{,} \PY{n}{Component}\PY{p}{,} \PY{n}{PhaseComponent}
\PY{k+kn}{import} \PY{n+nn}{pint}\PY{n+nn}{.}\PY{n+nn}{models}\PY{n+nn}{.}\PY{n+nn}{parameter} \PY{k}{as} \PY{n+nn}{p}


\PY{k}{class} \PY{n+nc}{PeriodSpindown}\PY{p}{(}\PY{n}{PhaseComponent}\PY{p}{)}\PY{p}{:}
    \PY{l+s+sd}{\PYZdq{}\PYZdq{}\PYZdq{}This is an example model component of pular spindown but parametrized}
\PY{l+s+sd}{    as period. }
\PY{l+s+sd}{    \PYZdq{}\PYZdq{}\PYZdq{}}

    \PY{n}{register} \PY{o}{=} \PY{k+kc}{True}  \PY{c+c1}{\PYZsh{} Flags for the model builder to find this component.}
    \PY{n}{category} \PY{o}{=} \PY{l+s+s2}{\PYZdq{}}\PY{l+s+s2}{spindown}\PY{l+s+s2}{\PYZdq{}} \PY{c+c1}{\PYZsh{} Give a category for the component sorting.}
    \PY{c+c1}{\PYZsh{} define the init function.}
    \PY{c+c1}{\PYZsh{} Most components do not have a parameter for input.}
    \PY{k}{def} \PY{n+nf+fm}{\PYZus{}\PYZus{}init\PYZus{}\PYZus{}}\PY{p}{(}\PY{n+nb+bp}{self}\PY{p}{)}\PY{p}{:}
        \PY{c+c1}{\PYZsh{} Get the attruibutes that initilzed in the parent class}
        \PY{n+nb}{super}\PY{p}{(}\PY{p}{)}\PY{o}{.}\PY{n+nf+fm}{\PYZus{}\PYZus{}init\PYZus{}\PYZus{}}\PY{p}{(}\PY{p}{)}
        \PY{c+c1}{\PYZsh{} Add parameters using the add\PYZus{}params in the TimingModel}
        \PY{c+c1}{\PYZsh{} Add spin period as parameter}
        \PY{n+nb+bp}{self}\PY{o}{.}\PY{n}{add\PYZus{}param}\PY{p}{(}
            \PY{n}{p}\PY{o}{.}\PY{n}{floatParameter}\PY{p}{(}
                \PY{n}{name}\PY{o}{=}\PY{l+s+s2}{\PYZdq{}}\PY{l+s+s2}{P0}\PY{l+s+s2}{\PYZdq{}}\PY{p}{,}
                \PY{n}{value}\PY{o}{=}\PY{k+kc}{None}\PY{p}{,}
                \PY{n}{units}\PY{o}{=}\PY{n}{u}\PY{o}{.}\PY{n}{s}\PY{p}{,}
                \PY{n}{description}\PY{o}{=}\PY{l+s+s2}{\PYZdq{}}\PY{l+s+s2}{Spin period}\PY{l+s+s2}{\PYZdq{}}\PY{p}{,}
                \PY{n}{longdouble}\PY{o}{=}\PY{k+kc}{True}\PY{p}{,}
            \PY{p}{)}
        \PY{p}{)}
        \PY{c+c1}{\PYZsh{} Add spin period derivative P1. Since it is not all rquired, we are setting the}
        \PY{c+c1}{\PYZsh{} default value to 0.0}
        \PY{n+nb+bp}{self}\PY{o}{.}\PY{n}{add\PYZus{}param}\PY{p}{(}
            \PY{n}{p}\PY{o}{.}\PY{n}{floatParameter}\PY{p}{(}
                \PY{n}{name}\PY{o}{=}\PY{l+s+s2}{\PYZdq{}}\PY{l+s+s2}{P1}\PY{l+s+s2}{\PYZdq{}}\PY{p}{,}
                \PY{n}{value}\PY{o}{=}\PY{l+m+mf}{0.0}\PY{p}{,}
                \PY{n}{units}\PY{o}{=}\PY{n}{u}\PY{o}{.}\PY{n}{s} \PY{o}{/} \PY{n}{u}\PY{o}{.}\PY{n}{s}\PY{p}{,}
                \PY{n}{description}\PY{o}{=}\PY{l+s+s2}{\PYZdq{}}\PY{l+s+s2}{Spin period derivative}\PY{l+s+s2}{\PYZdq{}}\PY{p}{,}
                \PY{n}{longdouble}\PY{o}{=}\PY{k+kc}{True}\PY{p}{,}
            \PY{p}{)}
        \PY{p}{)}
        \PY{c+c1}{\PYZsh{} Add reference epoch time.}
        \PY{n+nb+bp}{self}\PY{o}{.}\PY{n}{add\PYZus{}param}\PY{p}{(}
            \PY{n}{p}\PY{o}{.}\PY{n}{MJDParameter}\PY{p}{(}
                \PY{n}{name}\PY{o}{=}\PY{l+s+s2}{\PYZdq{}}\PY{l+s+s2}{PEPOCH\PYZus{}P0}\PY{l+s+s2}{\PYZdq{}}\PY{p}{,}
                \PY{n}{description}\PY{o}{=}\PY{l+s+s2}{\PYZdq{}}\PY{l+s+s2}{Reference epoch for spin\PYZhy{}down}\PY{l+s+s2}{\PYZdq{}}\PY{p}{,}
                \PY{n}{time\PYZus{}scale}\PY{o}{=}\PY{l+s+s2}{\PYZdq{}}\PY{l+s+s2}{tdb}\PY{l+s+s2}{\PYZdq{}}\PY{p}{,}
            \PY{p}{)}
        \PY{p}{)}
        \PY{c+c1}{\PYZsh{} Add spindown phase model function to phase functions}
        \PY{n+nb+bp}{self}\PY{o}{.}\PY{n}{phase\PYZus{}funcs\PYZus{}component} \PY{o}{+}\PY{o}{=} \PY{p}{[}\PY{n+nb+bp}{self}\PY{o}{.}\PY{n}{spindown\PYZus{}phase\PYZus{}period}\PY{p}{]}
        \PY{c+c1}{\PYZsh{} Add the d\PYZus{}phase\PYZus{}d\PYZus{}delay derivative to the list}
        \PY{n+nb+bp}{self}\PY{o}{.}\PY{n}{phase\PYZus{}derivs\PYZus{}wrt\PYZus{}delay} \PY{o}{+}\PY{o}{=} \PY{p}{[}\PY{n+nb+bp}{self}\PY{o}{.}\PY{n}{d\PYZus{}spindown\PYZus{}phase\PYZus{}period\PYZus{}d\PYZus{}delay}\PY{p}{]}
        \PY{c+c1}{\PYZsh{} Setup the unique parameters, if this component shares the same parameter}
        \PY{c+c1}{\PYZsh{} with other components. Otherwise this is optional.}
        \PY{n+nb+bp}{self}\PY{o}{.}\PY{n}{set\PYZus{}special\PYZus{}params}\PY{p}{(}\PY{p}{[}\PY{l+s+s1}{\PYZsq{}}\PY{l+s+s1}{P0}\PY{l+s+s1}{\PYZsq{}}\PY{p}{,} \PY{l+s+s1}{\PYZsq{}}\PY{l+s+s1}{P1}\PY{l+s+s1}{\PYZsq{}}\PY{p}{]}\PY{p}{)}

    \PY{k}{def} \PY{n+nf}{setup}\PY{p}{(}\PY{n+nb+bp}{self}\PY{p}{)}\PY{p}{:}
        \PY{l+s+sd}{\PYZdq{}\PYZdq{}\PYZdq{}Setup the model. Register the derivative functions\PYZdq{}\PYZdq{}\PYZdq{}}
        \PY{n+nb}{super}\PY{p}{(}\PY{p}{)}\PY{o}{.}\PY{n}{setup}\PY{p}{(}\PY{p}{)}  \PY{c+c1}{\PYZsh{} This will run the setup in the Component class.}
        \PY{c+c1}{\PYZsh{} The following lines are resgistering the derivative functions to the timingmodel.}
        \PY{n+nb+bp}{self}\PY{o}{.}\PY{n}{register\PYZus{}deriv\PYZus{}funcs}\PY{p}{(}\PY{n+nb+bp}{self}\PY{o}{.}\PY{n}{d\PYZus{}phase\PYZus{}d\PYZus{}P0}\PY{p}{,} \PY{l+s+s2}{\PYZdq{}}\PY{l+s+s2}{P0}\PY{l+s+s2}{\PYZdq{}}\PY{p}{)}
        \PY{n+nb+bp}{self}\PY{o}{.}\PY{n}{register\PYZus{}deriv\PYZus{}funcs}\PY{p}{(}\PY{n+nb+bp}{self}\PY{o}{.}\PY{n}{d\PYZus{}phase\PYZus{}d\PYZus{}P1}\PY{p}{,} \PY{l+s+s2}{\PYZdq{}}\PY{l+s+s2}{P1}\PY{l+s+s2}{\PYZdq{}}\PY{p}{)}

    \PY{k}{def} \PY{n+nf}{validate}\PY{p}{(}\PY{n+nb+bp}{self}\PY{p}{)}\PY{p}{:}
        \PY{l+s+sd}{\PYZdq{}\PYZdq{}\PYZdq{}Check the parameter value.\PYZdq{}\PYZdq{}\PYZdq{}}
        \PY{n+nb}{super}\PY{p}{(}\PY{p}{)}\PY{o}{.}\PY{n}{validate}\PY{p}{(}\PY{p}{)}  \PY{c+c1}{\PYZsh{} This will run the .validate() in the component class}
        \PY{c+c1}{\PYZsh{} Check required parameters, since P1 is not required, we are not checking it here}
        \PY{k}{for} \PY{n}{param} \PY{o+ow}{in} \PY{p}{[}\PY{l+s+s2}{\PYZdq{}}\PY{l+s+s2}{P0}\PY{l+s+s2}{\PYZdq{}}\PY{p}{]}\PY{p}{:}
            \PY{k}{if} \PY{n+nb}{getattr}\PY{p}{(}\PY{n+nb+bp}{self}\PY{p}{,} \PY{n}{param}\PY{p}{)} \PY{o+ow}{is} \PY{k+kc}{None}\PY{p}{:}
                \PY{k}{raise} \PY{n+ne}{ValueError}\PY{p}{(}\PY{l+s+s2}{\PYZdq{}}\PY{l+s+s2}{Spindown period model needs }\PY{l+s+si}{\PYZob{}\PYZcb{}}\PY{l+s+s2}{\PYZdq{}}\PY{o}{.}\PY{n}{format}\PY{p}{(}\PY{n}{param}\PY{p}{)}\PY{p}{)}

    \PY{c+c1}{\PYZsh{} One can always setup properties for updating attributes automatically.}
    \PY{n+nd}{@property}
    \PY{k}{def} \PY{n+nf}{F0}\PY{p}{(}\PY{n+nb+bp}{self}\PY{p}{)}\PY{p}{:}
        \PY{c+c1}{\PYZsh{} We return F0 as parameter here since the other place of PINT code use F0}
        \PY{c+c1}{\PYZsh{} in the format of PINT parameter.}
        \PY{k}{return} \PY{n}{p}\PY{o}{.}\PY{n}{floatParameter}\PY{p}{(}
            \PY{n}{name}\PY{o}{=}\PY{l+s+s2}{\PYZdq{}}\PY{l+s+s2}{F0}\PY{l+s+s2}{\PYZdq{}}\PY{p}{,}
            \PY{n}{value}\PY{o}{=}\PY{l+m+mf}{1.0} \PY{o}{/} \PY{n+nb+bp}{self}\PY{o}{.}\PY{n}{P0}\PY{o}{.}\PY{n}{quantity}\PY{p}{,}
            \PY{n}{units}\PY{o}{=}\PY{l+s+s2}{\PYZdq{}}\PY{l+s+s2}{Hz}\PY{l+s+s2}{\PYZdq{}}\PY{p}{,}
            \PY{n}{description}\PY{o}{=}\PY{l+s+s2}{\PYZdq{}}\PY{l+s+s2}{Spin\PYZhy{}frequency}\PY{l+s+s2}{\PYZdq{}}\PY{p}{,}
            \PY{n}{long\PYZus{}double}\PY{o}{=}\PY{k+kc}{True}\PY{p}{,}
        \PY{p}{)}

    \PY{c+c1}{\PYZsh{} Defining the derivatives. In the PINT, a common format of derivative naming is}
    \PY{c+c1}{\PYZsh{} d\PYZus{}xxx\PYZus{}d\PYZus{}xxxx}
    \PY{n+nd}{@property}
    \PY{k}{def} \PY{n+nf}{d\PYZus{}F0\PYZus{}d\PYZus{}P0}\PY{p}{(}\PY{n+nb+bp}{self}\PY{p}{)}\PY{p}{:}
        \PY{k}{return} \PY{o}{\PYZhy{}}\PY{l+m+mf}{1.0} \PY{o}{/} \PY{n+nb+bp}{self}\PY{o}{.}\PY{n}{P0}\PY{o}{.}\PY{n}{quantity} \PY{o}{*}\PY{o}{*} \PY{l+m+mi}{2}

    \PY{n+nd}{@property}
    \PY{k}{def} \PY{n+nf}{F1}\PY{p}{(}\PY{n+nb+bp}{self}\PY{p}{)}\PY{p}{:}
        \PY{k}{return} \PY{n}{p}\PY{o}{.}\PY{n}{floatParameter}\PY{p}{(}
            \PY{n}{name}\PY{o}{=}\PY{l+s+s2}{\PYZdq{}}\PY{l+s+s2}{F1}\PY{l+s+s2}{\PYZdq{}}\PY{p}{,}
            \PY{n}{value}\PY{o}{=}\PY{n+nb+bp}{self}\PY{o}{.}\PY{n}{d\PYZus{}F0\PYZus{}d\PYZus{}P0} \PY{o}{*} \PY{n+nb+bp}{self}\PY{o}{.}\PY{n}{P1}\PY{o}{.}\PY{n}{quantity}\PY{p}{,}
            \PY{n}{units}\PY{o}{=}\PY{n}{u}\PY{o}{.}\PY{n}{Hz} \PY{o}{/} \PY{n}{u}\PY{o}{.}\PY{n}{s}\PY{p}{,}
            \PY{n}{description}\PY{o}{=}\PY{l+s+s2}{\PYZdq{}}\PY{l+s+s2}{Spin down frequency}\PY{l+s+s2}{\PYZdq{}}\PY{p}{,}
            \PY{n}{long\PYZus{}double}\PY{o}{=}\PY{k+kc}{True}\PY{p}{,}
        \PY{p}{)}

    \PY{n+nd}{@property}
    \PY{k}{def} \PY{n+nf}{d\PYZus{}F1\PYZus{}d\PYZus{}P0}\PY{p}{(}\PY{n+nb+bp}{self}\PY{p}{)}\PY{p}{:}
        \PY{k}{return} \PY{n+nb+bp}{self}\PY{o}{.}\PY{n}{P1}\PY{o}{.}\PY{n}{quantity} \PY{o}{*} \PY{l+m+mf}{2.0} \PY{o}{/} \PY{n+nb+bp}{self}\PY{o}{.}\PY{n}{P0}\PY{o}{.}\PY{n}{quantity} \PY{o}{*}\PY{o}{*} \PY{l+m+mi}{3}

    \PY{n+nd}{@property}
    \PY{k}{def} \PY{n+nf}{d\PYZus{}F1\PYZus{}d\PYZus{}P1}\PY{p}{(}\PY{n+nb+bp}{self}\PY{p}{)}\PY{p}{:}
        \PY{k}{return} \PY{n+nb+bp}{self}\PY{o}{.}\PY{n}{d\PYZus{}F0\PYZus{}d\PYZus{}P0}

    \PY{k}{def} \PY{n+nf}{get\PYZus{}dt}\PY{p}{(}\PY{n+nb+bp}{self}\PY{p}{,} \PY{n}{toas}\PY{p}{,} \PY{n}{delay}\PY{p}{)}\PY{p}{:}
        \PY{l+s+sd}{\PYZdq{}\PYZdq{}\PYZdq{}dt from the toas to the reference time.\PYZdq{}\PYZdq{}\PYZdq{}}
        \PY{c+c1}{\PYZsh{} toas.table[\PYZsq{}tdbld\PYZsq{}] stores the tdb time in longdouble.}
        \PY{k}{return} \PY{p}{(}\PY{n}{toas}\PY{o}{.}\PY{n}{table}\PY{p}{[}\PY{l+s+s2}{\PYZdq{}}\PY{l+s+s2}{tdbld}\PY{l+s+s2}{\PYZdq{}}\PY{p}{]} \PY{o}{\PYZhy{}} \PY{n+nb+bp}{self}\PY{o}{.}\PY{n}{PEPOCH\PYZus{}P0}\PY{o}{.}\PY{n}{value}\PY{p}{)} \PY{o}{*} \PY{n}{u}\PY{o}{.}\PY{n}{day} \PY{o}{\PYZhy{}} \PY{n}{delay}

    \PY{c+c1}{\PYZsh{} Defining the phase function, which is added to the self.phase\PYZus{}funcs\PYZus{}component}
    \PY{k}{def} \PY{n+nf}{spindown\PYZus{}phase\PYZus{}period}\PY{p}{(}\PY{n+nb+bp}{self}\PY{p}{,} \PY{n}{toas}\PY{p}{,} \PY{n}{delay}\PY{p}{)}\PY{p}{:}
        \PY{l+s+sd}{\PYZdq{}\PYZdq{}\PYZdq{}Spindown phase using P0 and P1\PYZdq{}\PYZdq{}\PYZdq{}}
        \PY{n}{dt} \PY{o}{=} \PY{n+nb+bp}{self}\PY{o}{.}\PY{n}{get\PYZus{}dt}\PY{p}{(}\PY{n}{toas}\PY{p}{,} \PY{n}{delay}\PY{p}{)}
        \PY{k}{return} \PY{n+nb+bp}{self}\PY{o}{.}\PY{n}{F0}\PY{o}{.}\PY{n}{quantity} \PY{o}{*} \PY{n}{dt} \PY{o}{+} \PY{l+m+mf}{0.5} \PY{o}{*} \PY{n+nb+bp}{self}\PY{o}{.}\PY{n}{F1}\PY{o}{.}\PY{n}{quantity} \PY{o}{*} \PY{n}{dt} \PY{o}{*}\PY{o}{*} \PY{l+m+mi}{2}

    \PY{k}{def} \PY{n+nf}{d\PYZus{}spindown\PYZus{}phase\PYZus{}period\PYZus{}d\PYZus{}delay}\PY{p}{(}\PY{n+nb+bp}{self}\PY{p}{,} \PY{n}{toas}\PY{p}{,} \PY{n}{delay}\PY{p}{)}\PY{p}{:}
        \PY{l+s+sd}{\PYZdq{}\PYZdq{}\PYZdq{}This is part of the derivative chain for the parameters in the delay term.}
\PY{l+s+sd}{        \PYZdq{}\PYZdq{}\PYZdq{}}
        \PY{n}{dt} \PY{o}{=} \PY{n+nb+bp}{self}\PY{o}{.}\PY{n}{get\PYZus{}dt}\PY{p}{(}\PY{n}{toas}\PY{p}{,} \PY{n}{delay}\PY{p}{)}
        \PY{k}{return} \PY{o}{\PYZhy{}}\PY{p}{(}\PY{n+nb+bp}{self}\PY{o}{.}\PY{n}{F0}\PY{o}{.}\PY{n}{quantity} \PY{o}{+} \PY{n}{dt} \PY{o}{*} \PY{n+nb+bp}{self}\PY{o}{.}\PY{n}{F1}\PY{o}{.}\PY{n}{quantity}\PY{p}{)}

    \PY{k}{def} \PY{n+nf}{d\PYZus{}phase\PYZus{}d\PYZus{}P0}\PY{p}{(}\PY{n+nb+bp}{self}\PY{p}{,} \PY{n}{toas}\PY{p}{,} \PY{n}{param}\PY{p}{,} \PY{n}{delay}\PY{p}{)}\PY{p}{:}
        \PY{n}{dt} \PY{o}{=} \PY{n+nb+bp}{self}\PY{o}{.}\PY{n}{get\PYZus{}dt}\PY{p}{(}\PY{n}{toas}\PY{p}{,} \PY{n}{delay}\PY{p}{)}
        \PY{k}{return} \PY{n+nb+bp}{self}\PY{o}{.}\PY{n}{d\PYZus{}F0\PYZus{}d\PYZus{}P0} \PY{o}{*} \PY{n}{dt} \PY{o}{+} \PY{l+m+mf}{0.5} \PY{o}{*} \PY{n+nb+bp}{self}\PY{o}{.}\PY{n}{d\PYZus{}F1\PYZus{}d\PYZus{}P0} \PY{o}{*} \PY{n}{dt} \PY{o}{*}\PY{o}{*} \PY{l+m+mi}{2}

    \PY{k}{def} \PY{n+nf}{d\PYZus{}phase\PYZus{}d\PYZus{}P1}\PY{p}{(}\PY{n+nb+bp}{self}\PY{p}{,} \PY{n}{toas}\PY{p}{,} \PY{n}{param}\PY{p}{,} \PY{n}{delay}\PY{p}{)}\PY{p}{:}
        \PY{n}{dt} \PY{o}{=} \PY{n+nb+bp}{self}\PY{o}{.}\PY{n}{get\PYZus{}dt}\PY{p}{(}\PY{n}{toas}\PY{p}{,} \PY{n}{delay}\PY{p}{)}
        \PY{k}{return} \PY{l+m+mf}{0.5} \PY{o}{*} \PY{n+nb+bp}{self}\PY{o}{.}\PY{n}{d\PYZus{}F1\PYZus{}d\PYZus{}P1} \PY{o}{*} \PY{n}{dt} \PY{o}{*}\PY{o}{*} \PY{l+m+mi}{2}
\end{Verbatim}